\documentclass[%
 aip,
 jcp,% Journal of Chemical Physics
 amsmath,amssymb,
 preprint,% AIP formats peer-review/production under preprint; switch to reprint
]{revtex4-1}

\usepackage{graphicx}% Include figure files
\usepackage{dcolumn}% Align table columns on decimal point
\usepackage{bm}% bold math
\usepackage{longtable}% page-breaking grid table (App. B)

\usepackage[utf8]{inputenc}
\usepackage[T1]{fontenc}
\usepackage{mathptmx}
\usepackage{etoolbox}
\usepackage{placeins}% \FloatBarrier: keep figures within their section (out of Conclusions)
\makeatletter
\def\@email#1#2{%
 \endgroup
 \patchcmd{\titleblock@produce}
  {\frontmatter@RRAPformat}
  {\frontmatter@RRAPformat{\produce@RRAP{*#1\href{mailto:#2}{#2}}}\frontmatter@RRAPformat}
  {}{}
}%
\makeatother

\newcommand{\kB}{k_{B}}
\newcommand{\Scage}{\Delta S_{\rm cage}}
\newcommand{\FK}{F_{K}}
\newcommand{\FM}{F_{M}}
\newcommand{\dpar}{d_{\parallel}}

\begin{document}

\preprint{AIP/123-QED}

\title[Constructing the memory cage]{An anharmonic liquid-entropy functional from the Mori--Zwanzig memory kernel}

\author{Ricardo Buarque}
\author{Mauro Gascon}
\affiliation{Program in Materials Science and Engineering, University of California San Diego,
La Jolla, California 92093, USA}
\author{Tod A. Pascal}
 \email{tpascal@ucsd.edu}
\affiliation{Program in Materials Science and Engineering, University of California San Diego,
La Jolla, California 92093, USA}
\affiliation{ATLAS Materials Physics Laboratory, Aiiso Yufeng Li Family Department of Nano and
Chemical Engineering, University of California San Diego, La Jolla, California 92093, USA}

\date{\today}

\begin{abstract}
We utilize the Mori--Zwanzig memory kernel to separate the density of states of a liquid into gas, cage, and harmonic-solid components. The cage entropy is computed self-consistently as the non-Markovian excess of the full velocity response over its Markovian counterpart, and is assigned an excluded-volume entropy reference rather than a harmonic one. Employing physically motivated constraints, the resulting Three-Phase Explicit Anharmonic Thermodynamics (3PT) method reproduces thermodynamic-integration entropies of liquid metals and aligns two water models with independent free-energy perturbation benchmarks. For monoatomic Lennard-Jones liquids, 3PT's accuracy follows a universal efficiency curve governed by the kernel's non-Markovianity. We thus establish a direct, trajectory-level mapping between memory, transient cage dynamics, and liquid entropy.
\end{abstract}

\maketitle

%==============================================================
\section{Introduction}\label{sec:intro}
%==============================================================
The velocity autocorrelation function (VACF) is the central dynamical observable in density-of-states-based entropy estimators. Its cosine transform gives the spectral distribution of translational, rotational, or vibrational motion sampled by an equilibrium trajectory. This, in turn, leads naturally to a central question: once a spectrum has been obtained, how should the weights be assigned in order to obtain the entropy?

The Two-Phase Thermodynamics (2PT) method of Lin, Blanco and Goddard\cite{Lin2003} provides one physically transparent closure to this question. It partitions the velocity density of states (vDOS - the Fourier transform of the VACF) into a diffusive gas component, represented by a Lorentzian fixed by the zero-frequency response, and a residual solid component, assigned a harmonic-oscillator entropy weight. This construction is efficient because the VACF of a liquid converges on picosecond time scales, and avoids path-dependent thermodynamic integration (TI) or $\lambda$-dependent free-energy perturbation (FEP). Its limitation is equally clear: all non-diffusive spectral weight not assigned to the gas is treated as harmonic, even when it arises from finite-memory, anharmonic cage motion.

The mis-assigned cage motion represents an intermediate regime, which we will show has a direct interpretation in the Mori--Zwanzig generalized Langevin equation\cite{Mori1965,Zwanzig2001} (GLE). A tagged particle in a dense-liquid neither diffuses freely nor vibrates in a static harmonic well. It is transiently trapped by its neighbors (i.e., the cage), stores and returns momentum, and eventually escapes as the local structure rearranges. Molecular liquids add an analogous rotational channel: hindered librations of a water molecule, for example, are neither free rotations nor normal-mode vibrations over the full librational band.

Here we make that cage contribution explicit. We invert the normalized VACF to obtain the memory kernel, compute the full velocity response, and subtract the Markovian response at the same static friction. We then construct the cage spectrum over the low-frequency band, which is assigned an excluded-volume entropy reference rather than a harmonic-oscillator weight. Using several fixed, physically motivated constraints and no empirically fitted parameters, the resulting Three-Phase Explicit Anharmonic Thermodynamics (3PT) functional,
\(S_{\rm 3PT}=S_{\rm gas}+S_{\rm cage}+S_{\rm solid}\), is shown to recover the total entropy in proportion to the kernel non-Markovianity. For strongly caged liquid metals, it reproduces entropies calculated by TI. For liquid water, which is strongly caged in its rotational channel but only moderately caged in its translational channel, it predicts entropies in close agreement with independent FEP benchmarks. For weakly non-Markovian monoatomic Lennard-Jones liquids, it provides only a partial correction, which is a diagnostic feature: the ratio $\gamma/\Omega_0$ defines an order parameter between the static friction and Einstein frequency, which predicts how much of the entropy deviation is carried by memory and can therefore be recovered by a cage term.

The remainder of the paper is organized as follows. In Sec.~\ref{sec:theory} we develop the 3PT kernel-based entropy functional and define its cage closure. In Sec.~\ref{sec:results} we benchmark 3PT against 2PT variants and independent entropy references for Lennard-Jones liquids, liquid metals, and water. In Sec.~\ref{sec:discussions} we analyze the various constraints that complete the model, and the thermodynamic consistency of the resulting entropy surface.

%==============================================================
\section{Theory and Methods}\label{sec:theory}
%==============================================================

\subsection{The 2PT closure of the Mori--Zwanzig kernel}\label{sec:2pt}
From an equilibrium molecular-dynamics trajectory, we form the mass-weighted VACF $C(t)=\sum_j m_j\langle \bm v_j(0)\!\cdot\!\bm v_j(t)\rangle$, and its cosine transform, the density of states (DoS):
\begin{equation}
S(\nu)=\frac{4}{\kB T}\int_0^\infty C(t)\cos(2\pi\nu t)\,dt,\qquad \int_0^\infty S(\nu)\,d\nu = dN,
\end{equation}
normalized to the $d$ translational degrees of freedom per particle ($d$ is the spatial dimension). Molecular fluids carry separate translational, rotational, and vibrational channels; we develop the translational channel here and address rotations in Sec.~\ref{sec:water}. Note: to avoid ambiguity, we use $S(\nu)$ for the spectral density of states and $S$ without an argument for thermodynamic entropy.

The autocorrelation $C(t)$ is propagated by the GLE memory kernel $K(t)$ (Sec.~\ref{sec:cage}), with 2PT being a closure that represents its two limiting behaviors: a zero-memory (Markovian) friction and an infinite-stiffness harmonic restoring term. 2PT self-consistently partitions the resulting DoS into a gas and a solid component, $S(\nu)=S^{\rm gas}(\nu)+S^{\rm sol}(\nu)$. The gas component is a Lorentzian pinned to the zero-frequency intensity, $S^{\rm gas}(0)=S(0)$, whose integrated weight $dfN$ defines the fluidicity $f\in[0,1]$. This gas component is then assigned the entropy of a hard-sphere fluid at the corresponding packing fraction $y(f)$, with the per-mode weight:
\begin{equation}
W_g=\frac1d\Big[\tfrac d2+1+\ln\!\big(\Lambda^{-d}\,v/f\big)\Big]+\frac1d\,s^{\rm ex}_{\rm HS}(y),
\end{equation}
representing the Sackur--Tetrode translational entropy plus the Carnahan--Starling excess $s^{\rm ex}_{\rm HS}$, with $\Lambda$ being the thermal wavelength and $v$ the volume per particle. We denote this the 2PT-rigorous hard-sphere weight throughout, i.e., without the empirical $+\ln Z$ compressibility term of the original formulation\cite{Lin2003}. The solid component is assigned the quantum harmonic-oscillator entropy, with per-mode weight $W_s(\nu)=u/(e^{u}-1)-\ln(1-e^{-u})$, $u=hc\nu/\kB T$, whose classical limit is $1-\ln u$. The 2PT entropy is then $S_{\rm 2PT}=\int[\,S^{\rm gas}W_g+S^{\rm sol}W_s\,]\,d\nu$. Its accuracy is strictly limited by the harmonic treatment of the solid remainder, which we now address.

\subsection{Constructing the Mori--Zwanzig cage}\label{sec:cage}
The harmonic weight $W_s$ is exact for small oscillations in a static well; it poorly describes a particle confined in a transient cage that itself rearranges. We isolate this cage motion through the GLE for velocity:
\begin{equation}
m\,\dot{\bm v}(t)=-\int_0^t K(t-t')\,\bm v(t')\,dt'+\bm R(t),
\end{equation}
where the memory kernel $K(t)$ encodes the friction exerted by the cage, and $\bm R(t)$ is the projected random force. Both the normalized VACF $c(t)=C(t)/C(0)$ and the kernel obey the Volterra equation\cite{Brunner2017},
\begin{equation}
\dot{c}(t)=-\int_0^t K(t-t')\,c(t')\,dt'
\end{equation}
which we invert in the time domain to obtain $K(t)$ (see derivation in Section ~\ref{SI-app:derivation} of Supporting Information - SI). Its one-sided Fourier transform $\tilde K(\omega=2\pi c\nu)$ defines the Markovian friction $\gamma\equiv\tilde K(0)$. 

Two spectral responses bracket the dynamics: the full, memory-retaining response is $\FK(\nu)={\rm Re}[1/(i\omega+\tilde K(\omega))]$, while its Markovian (memoryless) counterpart, obtained by freezing $\tilde K(\omega)$ at its zero-frequency value $\gamma$, is the Lorentzian $\FM(\nu)=\gamma/(\gamma^2+\omega^2)$. $\FK$ is thermodynamically motivated and, by the fluctuation--dissipation theorem, represents the measured DoS reconstructed from the kernel. $\FM$ acts as the gas-like response, representing a purely diffusive particle at the same static friction. 

The cage spectrum is then defined as the excess where the memory response departs from memory-less pure diffusion:
\begin{equation}
\mathrm{cage}(\nu)=\mathrm{clip}\!\big[\,\kappa\,(\FK(\nu)-\FM(\nu)),\;0,\;S^{\rm sol}_2(\nu)\,\big],
\label{eq:cage-excess}
\end{equation}
bounding the cage by the available solid weight $S^{\rm sol}_2=\max(S-S^{\rm gas},0)$, where $\kappa=S(0)/\!\int c(t)\,dt$ is the normalization constant in DoS units. In practice, we replace the hard $\mathrm{clip}$/$\max$ operations with softplus--LogSumExp smoothing. $\FK$ and $\FM$ share the $1/\gamma$ limit and both integrate to $\tfrac{\pi}{2}c(0)$ over $\omega\in(0,\infty)$. Thus Eq.~\eqref{eq:cage-excess} is a zero-sum spectral redistribution: $\int_0^\infty\big(\FK-\FM\big)\,d\omega=0$, exactly where the cage stores and returns momentum, $\FK>\FM$ (i.e., the positive band the clip retains). The compensating remainder lies at the frequencies from which the trapped weight arises, which the clip safely removes.

\subsection{The cage entropy: hard-sphere, not harmonic}\label{sec:cageentropy}
The central thesis of this work is that cage modes should be weighted as a bounded hard-sphere fluid, not as harmonic oscillators. This formal ansatz (i.e., assigning a macroscopic hard-sphere equation of state to bounded, high-frequency modes) is heavily motivated by liquid-state perturbation theory (e.g., Weeks--Chandler--Andersen), which demonstrate that excluded volume, rather than soft attractive interactions, dominates the configurational entropy\cite{WCA1971}. Thus, we replace the harmonic weight $W_s$ with the hard-sphere weight $W_g$ for the cage fraction, yielding the correction:
\begin{equation}\label{eq:dScage}
\Scage=p\,g(f)\int \mathrm{cage}(\nu)\,\big[1-w(\nu)\big]\,\big[\,W_g-W_s(\nu)\,\big]\,d\nu.
\end{equation}
Here, the spectrum, friction, Einstein frequency $\Omega_0$, and fluidicity are read directly from the trajectory, while the prefactor $p$ is constrained by dimensional consistency (Sec.~\ref{sec:theorem}). Finally, the gate $g(f)$ (see below) and filter $1-w(\nu)$ are physically motivated structural constraints introduced to enforce exact limiting behavior, rather than system-specific empirical tuning. The total entropy remains an exact sum over the gas, cage, and harmonic-solid components:
\begin{equation}
S_{\rm 3PT}=\underbrace{\textstyle\int S^{\rm gas}\,W_g\,d\nu}_{S_{\rm gas}}
+\underbrace{\textstyle\int \mathrm{cage}\,\big[\,W_s+p\,g(f)(1-w)(W_g-W_s)\,\big]d\nu}_{S_{\rm cage}}
+\underbrace{\textstyle\int S^{\rm sol}\,W_s\,d\nu}_{S_{\rm solid}},
\end{equation}
with $S=S^{\rm gas}+S^\mathrm{cage}+S^{\rm sol}$. We elaborate on the sensitivity of our calculated entropies to these constraints in Section ~\ref{SI-app:scope}.

A consistent entropy functional must smoothly interpolate between exact end-points. Therefore, the cage must vanish both in the dilute hard-sphere gas limit ($f\to1$ -- automatically satisfied as the available solid weight approaches zero) and in the harmonic crystal limit ($f\to0$). The latter is not automatically satisfied: a phonon spectrum intrinsically departs from a Markovian Lorentzian, meaning $\FK-\FM$ would spuriously over-correct a solid if left unchecked. Thus, we introduce a structural fluidicity gate:
\begin{equation}
g(f)=\frac{f^2}{f^2+f_0^2},
\end{equation}
which strictly vanishes as $f\to0$. Consequently, $\Scage\to0$ in the crystalline regime, returning 3PT precisely to the Debye vibrating crystal. The parameter $f_0$ acts as a structural gate rather than an empirical fit parameter; because homogeneous fluids naturally occupy $f\gtrsim0.15$ and solids $f\lesssim4\times10^{-4}$, any value of $f_0$ within this vast, empty gap produces identical results.

\subsection{Dimensional closure for the cage prefactor, $p=1/d$}\label{sec:theorem}
The prefactor $p$ in Eq.~\eqref{eq:dScage} scales the cage entropy, and is set by a dimensional-consistency counting. Since the memory kernel and the spectral responses ($\FK,\FM$) are inverted from the trace VACF $c(t)$, they are intrinsically normalized per-degree-of-freedom. An isotropic molecular cage has an interaction potential $U=U(r)$, and in the configurational integral, the angular surface $\Omega_{d-1}$ is identical for the anharmonic cage and its harmonic reference, canceling precisely in the excess. The total anharmonic excess is therefore carried by the single radial coordinate ($\dpar=1$), even though the velocity spectrum intrinsically superposes all $d$ channels. Summing this per-coordinate excess over the $\dpar$ coordinates that carry it gives:
\begin{equation}
\Scage=\dpar\,\mathcal S[\mathrm{cage}_\perp]=\frac{\dpar}{d}\,\mathcal S[\mathrm{cage}],
\qquad p=\frac{\dpar}{d},
\label{eq:pdd}
\end{equation}
meaning that an isotropic, angularly stiff liquid cage has $p=1/d$. In other words, $p$ is a counting factor derived from the dimensionality of the velocity projection, not a spectral memory weight. For a more complete derivation, we refer the interested reader to Sec.~\ref{SI-app:derivation}.

\subsection{Simulation protocol}\label{sec:protocol}
We generated all trajectories from equilibrium molecular dynamics (MD) simulations. For the 2D/3D systems we used LAMMPS~\cite{Plimpton1995,Thompson2022}, removing the system's center-of-mass (COM)
drift so that the velocity DoS is free of the spurious zero-frequency weight that COM motion
would contribute. The time integration closely followed the time-reversible, measure-preserving Verlet integrators of Tuckerman \textit{et al.}\cite{Tuckerman2006}. A 4D LJ system was simulated using a custom, in-house code (see below).

\paragraph{Lennard-Jones.} For the LJ(12-6) system we used argon parameters ($\sigma=3.405$~\AA,
$\varepsilon/\kB=119.78$~K, $m=39.948$~g\,mol$^{-1}$) with $N=512$ atoms (simple-cubic initial
lattice) for the fluid states and $N=500$ (FCC) for the solid, in the canonical (NVT) ensemble with a Nos\'e--Hoover thermostat (3 chains, relaxation time $\tau=50$~fs). We used a time step of 8~fs, equilibrated each state for 400~ps, and collected 250~ps of production, writing velocities every step (i.e., a $8$~fs dump interval) to give 31\,251 frames, fine
enough to resolve the fast initial decay of the memory kernel on which the cage entropy depends (Sec.~\ref{sec:lawnm}). Comparative simulations using a coarser $64$-fs trajectory dump sampling biased the cage construction (Fig.~\ref{fig:lawnm}). We treated the LJ dispersion by particle-mesh Ewald (\texttt{pair\_style lj/long}, \texttt{kspace\_style pppm/disp})
beyond a $5\sigma$ short-range cutoff rather than by an analytical tail correction. This ensured that the
forces, and hence the inverted memory kernel $K(t)$, are free of the cutoff artifact that
an energy/pressure-only tail correction would leave. Reference entropies were calculated using the modified
Benedict--Webb--Rubin (MBWR) equation of state of Johnson, Zollweg, and Gubbins~\cite{Johnson1993}
for the fluid and the van der Hoef
equation of state~\cite{vdHoef2000} for the FCC solid. The full $(\rho^*,T^*)$ grid is presented in
Table~\ref{SI-tab:s1}.

\paragraph{Four-dimensional Lennard-Jones.} A cross-dimensional test of the
prediction $p=1/4$ requires a genuinely four-dimensional liquid, which LAMMPS and other standard MD
codes (hard-wired to three Cartesian components) do not natively support. Thus, following the approach of Hloucha and Sandler~\cite{Hloucha1999}, we
created a 4D MD engine that integrates the LJ(12-6) potential on a
four-torus hypersurface with a force-shifted cutoff at $r_c=2.5\,\sigma$, Nos\'e--Hoover NVT (and an optional
Berendsen barostat), brute-forced $O(N^2)$ forces, initialized on a $D_4$ or hypercubic lattice. We
ran $N=6^4=1296$ atoms with an $8$-fs-equivalent step (argon reduced units) and velocities written
every step. The decisive state is $\rho^*=1.10$, $T^*=1.0$, which is strongly caged
($\gamma/\Omega_0=3.2$, on the saturated plateau of Sec.~\ref{sec:lawnm}); a $T^*=3.0$ and a
(supercritical) $T^*=4.0$ isotherm spanning $\rho^*=0.01$--$2.3$ and a $P^*=3.2$ isobar map the fluid
(the $4d$ critical point is $T_c^*\!\approx\!3.40$, $\rho_c^*\!\approx\!0.34$~\cite{Hloucha1999}). To establish the ground truth, we performed a
thermodynamic-integration (TI) reference calculation from the simulated equation of state. We used a low-density
virial anchor plus $\int(Z-1)/\rho\,d\rho$ from the ideal gas for the fluid branch and harmonic /
Frenkel--Ladd lattice dynamics for the $D_4$ crystal. We employed the 4D Sackur--Tetrode
ideal-gas term, and the 4D hard-sphere gas weight used the Luban--Michels equation of state
(virial coefficients $b_2=8$, $b_3=32.41$, $b_4=77.75$), which (with close packing at $\eta_{\rm cp}=\pi^2/16$) was validated against published $d=4$ Monte-Carlo simulations~\cite{SantosYusteLopezdeHaro2020}. We analyzed the trajectories with the same framework as the 2D and 3D systems: the velocity DoS, fluidicity, and gas/solid partition were generalized to arbitrary $d$ (the diffusive-Lorentzian half-width $2dNf$, the hard-sphere DoF
integral, the Sackur--Tetrode and hard-sphere-excess weights, and the active-DoF count all carry $d$
explicitly), with the memory-cage construction of Sec.~\ref{sec:theory} which is already dimension-agnostic.

\paragraph{Liquid metals.} We ran equilibrium MD simulations in LAMMPS for seven metals (Al, Ag, Ni, Pd, Rh, Au, Ir) with the
embedded-atom (\texttt{eam/fs}) Sutton--Chen potentials\cite{Sutton1990}, $N=512$ atoms (FCC $4\times4\times8$). We reproduced the fixed-density protocol of Sun \textit{et al.}~\cite{Sun2017}: a 1~fs time step, NVT at the
experimental density at melting, melting at high $T$ then cooling to the target $T$,
$\sim50$~ps of equilibration, and 100~ps of production with velocities
dumped every 10~fs. Here, our reference is the TI ionic entropy reported by Sun \textit{et al.}~\cite{Sun2017}.

\paragraph{Water.} We ran rigid TIP4P/2005~\cite{AbascalVega2005} with $N=1728$ molecules and a
1~fs time step, enforcing the rigid geometry by SHAKE constraints and treating the electrostatics by the \texttt{pppm/tip4p} method\cite{hockney2021}. We equilibrated the box under constant-pressure (1 atm), constant-temperature (298~K; NPT) dynamics with the Andersen barostat (pressure relaxation constant of 1~ps), using the equations of motion of Shinoda \textit{et al.}\cite{Shinoda2004}, which combine the hydrostatic equations of Martyna \textit{et al.}\cite{Martyna1994} with the strain energy of Parrinello and Rahman\cite{Parrinello1981}. This was followed by NVT production at 298~K. We applied the cage correction to
both the translational and the rotational channel (the latter with the free-rotor gas weight and
$p_{\rm rot}=1/3$; detailed in Sec.~\ref{sec:water}), with rotational symmetry number 2.
Our independent reference entropies are our previously published 2PT values~\cite{Pascal2011}, a Frenkel--Ladd /
Einstein-crystal\cite{FrenkelLadd1984,frenkel_smit_2002} absolute entropy, and a self-solvation excess-chemical-potential ($\mu_{\rm ex}$) FEP route (two-leg deletion with the Bennett acceptance-ratio estimator; both computed in this work,
Sec.~\ref{sec:water}), together with the experimental CODATA\cite{Mohr2005} value. We repeated the analysis for rigid SPC/E
water~\cite{Berendsen1987} (pppm electrostatics, $N=1728$ molecules), whose self-solvation free energy is
independently established in the
literature~\cite{WeberMerchantAsthagiri2011,MerchantAsthagiri2011,ShirtsPande2005}. We obtained the standard
molar entropy from $\mu_{\rm ex}$ with the residual-pressure standard-state term
(Sec.~\ref{sec:water}).

\paragraph{Finite-size corrections.}
For finite simulation cells, the long-time hydrodynamic tail of the
VACF lowers the self-diffusivity.
We thus apply the Yeh--Hummer corrections\cite{Yeh2004},
\(D_{\infty}=D_{\rm PBC}+\xi_T k_B T/(6\pi \eta L)\), with
\(\xi_T=2.837297\) for a cubic box. Presently, in 2PT/3PT, this
correction enters only through the diffusivity and hence the fluidicity. We do not modify the measured DoS, the spectral filter, or the
memory-derived cage term directly, because the Yeh--Hummer correction is
a long-time / low-frequency hydrodynamic effect. Unless otherwise noted,
we do not apply FSC corrections to the benchmarks reported in Sec.~\ref{sec:results}, so that
all methods are compared on the same trajectory-only basis.

\paragraph{Analysis.} All entropies were computed with our Python-based \texttt{py-xPT}\cite{py-xPT_2026_commit} code from the mass-weighted velocity
autocorrelation function (VACF via FFT, Hann apodization), partitioned as in
Secs.~\ref{sec:2pt}--\ref{sec:cageentropy} with the rigorous hard-sphere weight
(Carnahan--Starling EOS, no $+\ln Z$). All reported entropies are the
quantum-weighted values ($S_q$ in the 2PT literature). The cage post-correction of
Secs.~\ref{sec:cageentropy}--\ref{sec:theorem} ($p=1/d$) adds $\Scage$. We also calculated the thermodynamics of the 2PT variants from the identical trajectories. The entropy of water, being a molecular liquid, was decomposed into translational and rotational channels (the additional vibrational channel is zero by construction for the rigid TIP4P/2005 and SPC/E water models); monoatomic systems carry the translational channel only. For representative states we estimated the uncertainty in the reported entropies by block averaging / bootstrap over independent VACF segments; corresponding FEP/TI uncertainties are reported where available.

%==============================================================
\section{Results}\label{sec:results}
%==============================================================
The benchmark is identical across material classes: we compare our parameter-free 3PT entropy functional
against the two empirically corrected 2PT methods (the canonical version of Lin\cite{Lin2003} and the memory function DoS variant of Desjarlais\cite{Desjarlais2013}, both with the $+\ln Z$ compressibility
term), the R2PT reparameterization of Sun\cite{Sun2017}, and various independent references: the MBWR equation
of state for LJ, TI for metals, and FEP/TI for water. Rigorous-HS 2PT without a correction is shown as the uncorrected baseline.

\subsection{Monoatomic Lennard-Jones fluids vs.\ MBWR}\label{sec:lj}
We report 3PT entropies (with $p=1/3$) across a dense
$(\rho^*,T^*)$ grid of 71 monoatomic-LJ state points, comparing against the MBWR reference and the four 2PT entropy functionals computed from the same
trajectories. Over the 43 homogeneous-liquid
points ($0.70\le\rho^*\le1.02$), 3PT reproduces MBWR with a root-mean-square deviation of
$0.16\,\kB$ (mean $-0.15$). Comparatively, 3PT removes roughly half of the $0.33\,\kB$ deficit of uncorrected
rigorous-HS (which systematically underpredicts the entropy), while the $+\ln Z$ family (Lin-2003 and Desjarlais, both $\sim\!+0.14$) overpredicts. In contrast, the tuned R2PT ($0.10\,\kB$) is closest to MBWR. Table~\ref{tab:lj}
lists five representative liquid points; the complete grid is in Table~\ref{SI-tab:s1}.

\begin{table}[t]
\caption{\label{tab:lj}Molar entropy $S^*$ ($\kB$/atom) of the LJ fluid at five representative liquid
state points: MBWR reference, uncorrected rigorous-HS, Lin-2003 ($+\ln Z$), Desjarlais, R2PT, and
3PT. Deviations from MBWR in parentheses (512 atoms; all values this work).}
\begin{ruledtabular}
\begin{tabular}{lcccccc}
$\rho^*$/$T^*$ & MBWR & rig-HS & Lin-2003 & Desj. & R2PT & 3PT \\
\colrule
0.85/0.9 & 6.898 & 6.569 ($-$0.33) & 7.037 ($+$0.14) & 7.055 ($+$0.16) & \textbf{6.811 ($-$0.09)} & 6.805 ($-$0.09) \\
0.85/1.1 & 7.420 & 7.068 ($-$0.35) & 7.547 ($+$0.13) & 7.566 ($+$0.15) & \textbf{7.347 ($-$0.07)} & 7.279 ($-$0.14) \\
0.85/1.4 & 7.992 & 7.677 ($-$0.32) & 8.165 ($+$0.17) & 8.187 ($+$0.19) & \textbf{7.983 ($-$0.01)} & 7.863 ($-$0.13) \\
0.70/1.1 & 8.487 & 8.124 ($-$0.36) & 8.617 ($+$0.13) & 8.637 ($+$0.15) & \textbf{8.427 ($-$0.06)} & 8.247 ($-$0.24) \\
0.70/1.4 & 8.990 & 8.675 ($-$0.32) & 9.161 ($+$0.17) & 9.181 ($+$0.19) & \textbf{8.981 ($-$0.01)} & 8.792 ($-$0.20) \\
\colrule
RMS (these 5) & --- & 0.34 & 0.15 & 0.17 & \textbf{0.06} & 0.17 \\
RMS (43 liquids) & --- & 0.33 & 0.14 & 0.16 & \textbf{0.10} & 0.16 \\
\end{tabular}
\end{ruledtabular}
\end{table}

\begin{figure}[t]
\includegraphics[width=\columnwidth]{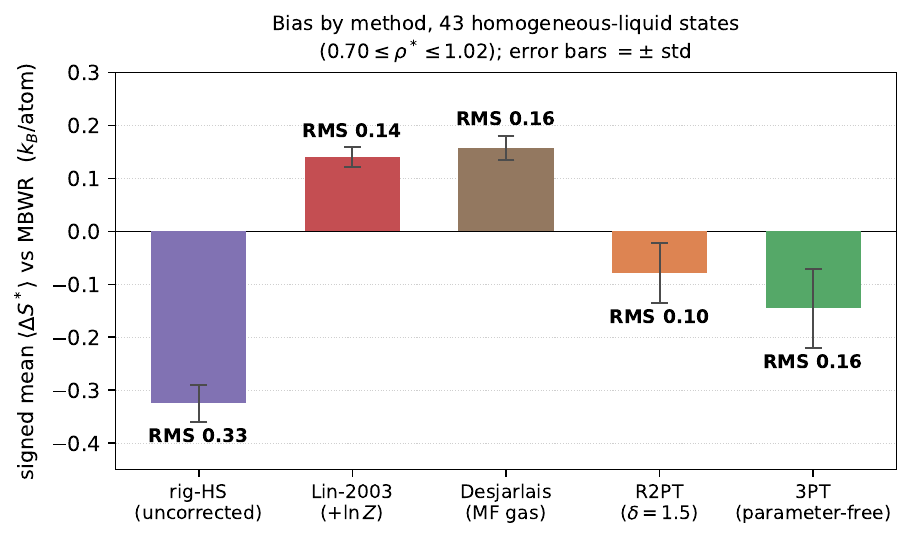}
\caption{\label{fig:bias}Per-method signed bias of the LJ entropy versus the MBWR reference,
averaged over the 43 homogeneous-liquid states ($0.70\le\rho^*\le1.02$); bars are the mean
deviation, error bars $\pm1\sigma$, RMS annotated. Uncorrected rigorous-HS runs low, the two
$+\ln Z$-convention methods (Lin-2003 and Desjarlais) overshoot by the same amount, and 3PT partially closes the gap, landing between them but short of the
system-tuned R2PT($\delta{=}1.5$), because the weak non-Markovian memory of monoatomic LJ
liquids limits how much of the deficit the explicit memory cage can recover (Sec.~\ref{sec:lawnm}).}
\end{figure}

\begin{figure}[t]
\includegraphics[width=\columnwidth]{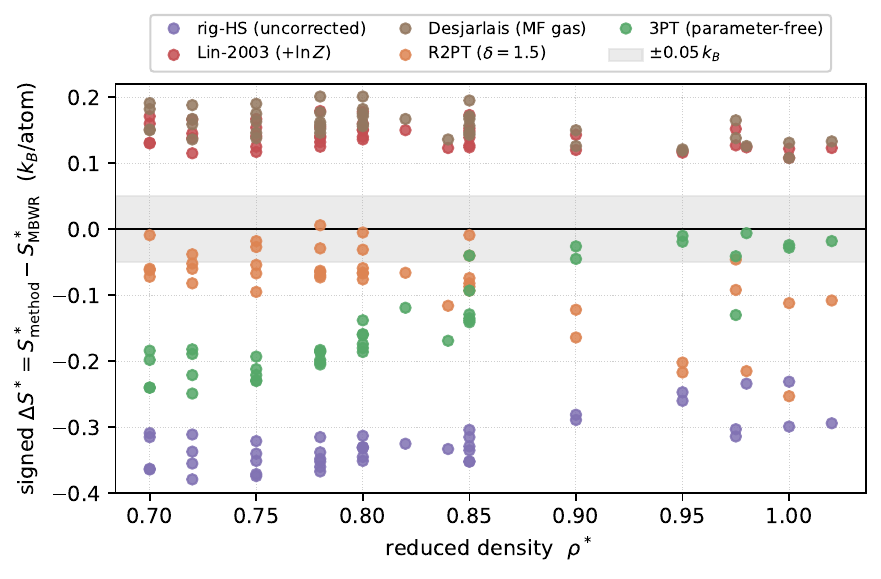}
\caption{\label{fig:signed}Per-state signed deviation $\Delta S^*=S^*_{\rm method}-S^*_{\rm MBWR}$
across the LJ liquid grid versus reduced density $\rho^*$. rig-HS (purple) sits systematically below
zero, Lin-2003 (red) and Desjarlais (brown) above (nearly coincident), and 3PT
(green) lies below zero by $\sim0.1$--$0.2\,\kB$, with the under-correction growing toward the warm,
lower-density edge where the memory cage is weakest, while the system-tuned R2PT($\delta{=}1.5$,
orange) stays closest to the reference.}
\end{figure}

The cage
correction in 3PT is confined to the intermediate liquid regime: it vanishes in the dilute-gas limit
and is explicitly gated off in the near-crystalline limit (gas and solid rows in
Table~\ref{SI-tab:s1}). Within the liquid set, the 3PT under-prediction is not uniform
(Fig.~\ref{fig:signed}): it is largest at the warm, lower-density edge ($\rho^*=0.70$, where 3PT
runs $0.20$--$0.24\,\kB$ low) but eases to $0.06$--$0.13\,\kB$ by
$\rho^*\!\approx\!0.85$--$1.0$. The origin is the memory content of the
kernel: the rigorous-HS deficit is nearly density-independent
($0.333\pm0.036\,\kB$ over the liquid grid), whereas the cage increment 3PT supplies grows with
the degree of caging, from $+0.13\,\kB$ at $\rho^*=0.70$ to $+0.27$--$0.28$ near
$\rho^*\!\approx\!1.0$. In the more fluid states bordering the supercritical region the
non-Markovian memory excess $\FK-\FM$ is small, so the cage offsets only part of a deficit whose magnitude is fixed by a largely memory-independent
anharmonicity.

\subsection{Liquid metals vs.\ thermodynamic integration}\label{sec:metals}
\begin{figure}[t]
\includegraphics[width=\columnwidth]{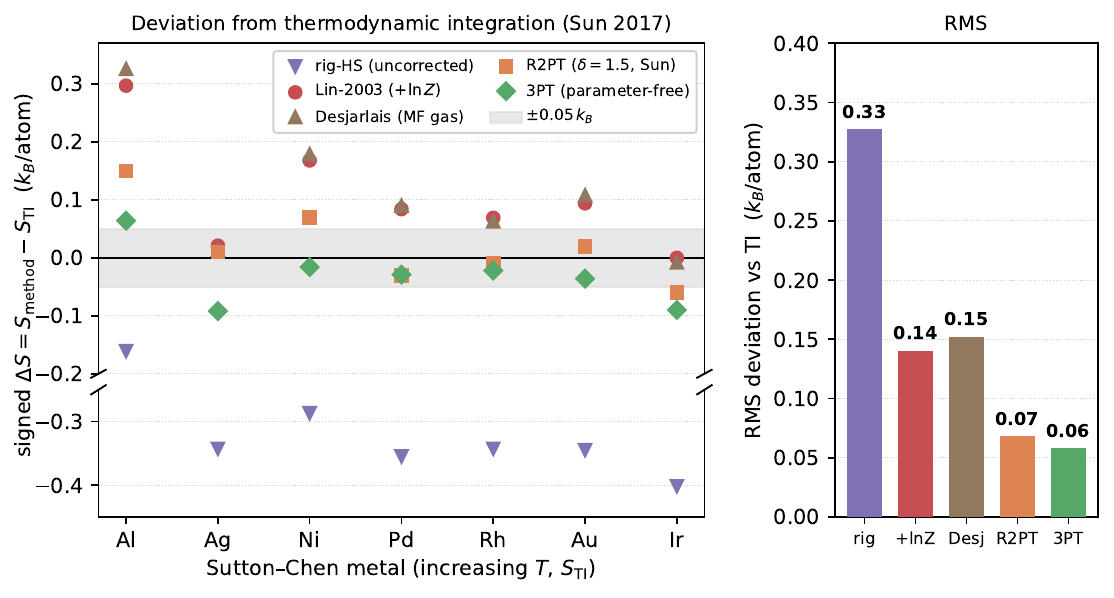}
\caption{\label{fig:metals}Deviation of seven Sutton--Chen liquid-metal entropies from the Sun
\textit{et al.}\ (2017) thermodynamic-integration reference, for all five entropy functionals
(left, per metal on a broken $y$-axis; right, RMS by method). rig-HS, Lin-2003 ($+\ln Z$),
Desjarlais, and 3PT are computed here from the same trajectories; R2PT($\delta=1.5$) is Sun's tuned
benchmark. 3PT is unbiased and matches the tuned R2PT; the $+\ln Z$ family (Lin-2003
and the nearly coincident Desjarlais) overestimates the entropy (worst at Al), and the uncorrected rig-HS
underestimates by $\sim0.3\,\kB$ (lower axis segment).}
\end{figure}

A decisive test of transferability is liquid metals, for which Sun \textit{et al.}\ \cite{Sun2017} computed
the entropy of seven Sutton--Chen metals using TI and developed the
tuned R2PT correction (an adjustable gas-fraction exponent $\delta$, optimized to
$\delta=1.5$) precisely because the canonical $+\ln Z$ 2PT overestimates metal entropies. As shown in Figure~\ref{fig:metals}, applying
3PT unchanged (same $p=1/3$ as LJ) reproduces the TI reference without any tuning
(Table~\ref{tab:metals}): the cage supplies a uniform $\Scage=+1.9$ to $+2.7$~J\,mol$^{-1}$K$^{-1}$
that lands the entropy within 1~$\sigma$ of TI, with a root-mean-square deviation of $0.06\,\kB$/atom and an unbiased mean of
$-0.03$, matching the accuracy of Sun's fitted R2PT ($\delta=1.5$, RMS $0.07$). In contrast, the canonical 2PT (RMS $0.14$) and the closely related Desjarlais moment-matched gas DoS
(RMS $0.15$) systematically overestimate the entropy, most severely for the soft-potential metal Al ($+0.30$ and $+0.33\,\kB$; $n=7$). Conversely, the uncorrected rigorous 2PT underestimates the entropy of these metals, with an RMS of $0.33$. Thus the failure modes that Sun's $\delta$-tuning
was built to repair are here recovered in 3PT due to the memory cage.

\begin{table}[t]
\caption{\label{tab:metals}Ionic entropy of seven Sutton--Chen liquid metals near melting
($\kB$/atom). The TI reference and the tuned R2PT($\delta=1.5$) are from Sun \textit{et al.}\ (2017),
Table~II; rig-HS, Lin-2003 ($+\ln Z$), Desjarlais, and 3PT are computed here from the same
trajectories (512 atoms). Deviations from TI in parentheses.}
\begin{ruledtabular}
\begin{tabular}{lccccccc}
Metal & $T$/K & TI (ref) & rig-HS & Lin-2003 & Desj. & R2PT(1.5)\footnote{From Sun \textit{et al.}\ (2017), Table~II.} & 3PT \\
\colrule
Al & 900 & 9.25 & 9.09 ($-$0.16) & 9.55 ($+$0.30) & 9.58 ($+$0.33) & 9.40 ($+$0.15) & \textbf{9.31 ($+$0.06)} \\
Ag & 1200 & 10.24 & 9.90 ($-$0.34) & 10.26 ($+$0.02) & 10.25 ($+$0.01) & 10.25 ($+$0.01) & \textbf{10.15 ($-$0.09)} \\
Ni & 1700 & 10.31 & 10.02 ($-$0.29) & 10.48 ($+$0.17) & 10.49 ($+$0.18) & 10.38 ($+$0.07) & \textbf{10.29 ($-$0.02)} \\
Pd & 1800 & 11.08 & 10.72 ($-$0.36) & 11.16 ($+$0.08) & 11.17 ($+$0.09) & 11.05 ($-$0.03) & \textbf{11.05 ($-$0.03)} \\
Rh & 2200 & 11.10 & 10.76 ($-$0.34) & 11.17 ($+$0.07) & 11.16 ($+$0.06) & 11.09 ($-$0.01) & \textbf{11.08 ($-$0.02)} \\
Au & 1300 & 11.71 & 11.37 ($-$0.35) & 11.80 ($+$0.09) & 11.82 ($+$0.11) & 11.73 ($+$0.02) & \textbf{11.67 ($-$0.04)} \\
Ir & 2700 & 12.25 & 11.85 ($-$0.40) & 12.25 ($+$0.00) & 12.24 ($-$0.01) & 12.19 ($-$0.06) & \textbf{12.16 ($-$0.09)} \\
\colrule
RMS dev. & & --- & 0.33 & 0.14 & 0.15 & 0.07 & \textbf{0.06} \\
\end{tabular}
\end{ruledtabular}
\end{table}

\subsection{Liquid water (TIP4P/2005 and SPC/E) vs Free Energy Perturbation}\label{sec:water}
Molecular liquids carry a rotational channel that is more caged than its translations: a water molecule neither
rotates freely nor librates harmonically but is a librational captive of its hydrogen-bonded
neighbors. We therefore apply the analogous cage construction to the angular-velocity autocorrelation function (derivation and normalization given in Sec.~\ref{SI-app:rot3pt}). 

For the rigid TIP4P/2005 at 298~K, the translational cage adds $+1.4$ and the rotational cage a further
$+5.2$~J\,mol$^{-1}$K$^{-1}$ to the rigorous-HS entropy ($54.6$); with a $+0.6$~J\,mol$^{-1}$K$^{-1}$
finite-size (Yeh--Hummer) correction this gives
$S^\circ_{\rm 3PT}=61.8$~J\,mol$^{-1}$K$^{-1}$ (Table~\ref{tab:water}).
The rotational increment reclassifies part of the hindered libration band ($\sim\!200$--$600$
cm$^{-1}$) from harmonic to caged free-rotor and is the larger of the two. This captures the right physics, since
libration is the more strongly hindered motion. The full translational+rotational 3PT entropy ($61.8$)
sits just $0.1$~J\,mol$^{-1}$K$^{-1}$ below our independent FEP self-solvation value
($61.9\pm0.5$~J\,mol$^{-1}$K$^{-1}$, this work; five-seed BAR), i.e.\ within its statistical uncertainty.
A residual systematic remains: since
$\partial S^\circ/\partial\langle U_{\rm inter}\rangle=1/T$, a $0.1$~kcal\,mol$^{-1}$ error in the
intermolecular energy alone shifts $S^\circ_{\rm FEP}$ by $\sim\!1.4$~J\,mol$^{-1}$K$^{-1}$, so we
treat this single-model agreement as a consistency check rather than a stringent test. The
translational cage alone ($56.0$) recovers only part of the deficit, so the rotational libration cage
is essential to a complete molecular entropy. A Frenkel--Ladd / Einstein-crystal absolute entropy ($67\pm5$, this work) and the
experimental $69.95$ sit above. The deviation from experiment reflects fundamental limitations of the water model: the expected classical-vs-quantum and rigid-model
bias.

\begin{table}[t] 
\caption{\label{tab:water}Standard molar entropy of liquid TIP4P/2005 water at 298~K
(J\,mol$^{-1}$K$^{-1}$). The translational cage
and the rotational (libration) cage together bring the rigorous-HS baseline onto the independent FEP
free-energy value.}
\begin{ruledtabular}
\begin{tabular}{lcl}
Method & $S^\circ_{\rm liq}$ & note \\
\colrule
rigorous-HS 2PT (uncorrected) & 54.6 & baseline, no correction \\
3PT ($+$ translational cage) & 56.0 & $\Scage^{\rm tr}=+1.4$ \\
canonical-2PT & 58.2 & empirical $+\ln Z$ \\
2PT, our previous work\cite{Pascal2011} & 57.5 & calculated using canonical-2PT \\
3PT (translational $+$ rotational cage) & \textbf{61.8} & $\Scage=+6.6$ ($+0.6$ Yeh--Hummer) \\
FEP self-solvation (this work) & $61.9\pm0.5$ & four-site BAR-FEP, $\mu_{\rm ex}=-7.19\pm0.04$ \\
Frenkel--Ladd / Einstein crystal (this work) & $67\pm5$ & absolute TI \\
experiment & 69.95 & CODATA\\
\end{tabular}
\end{ruledtabular}
\end{table}

\textit{The SPC/E water model, an independent free-energy anchor.} To test transferability across
water force fields, and to pin the molecular cage against a free-energy benchmark that is
independently established in the literature rather than computed only here, we repeated the
analysis for rigid SPC/E water~\cite{Berendsen1987} with the same $N=1728$ molecules. We find that the translational cage adds $+1.4$ and the
rotational libration cage a further $+5.0$~J\,mol$^{-1}$K$^{-1}$ to the rigorous-HS entropy ($57.1$);
with a $+0.5$~J\,mol$^{-1}$K$^{-1}$ Yeh--Hummer correction this
gives $S^\circ_{\rm 3PT}=64.0$~J\,mol$^{-1}$K$^{-1}$ (Table~\ref{tab:waterB}). Two features stand out.
First, both cage increments are transferable: the rotational (libration) increment ($+5.0$) is
essentially identical to that of TIP4P/2005 ($+5.2$), and, once the friction integral is taken to the
main lobe of the kernel rather than into its noisy tail (Sec.~\ref{SI-app:scope}), so is the
translational increment ($+1.4$, matching the $+1.4$ for TIP4P/2005). The cages are model-independent
features of liquid water, not artifacts of one parameterization. Second, the 3PT entropy
lands close to the formally exact FEP calculation, with a small translational residual. Our two-leg deletion BAR-FEP (five seeds) yields an excess chemical potential
$\mu_{\rm ex}$ of a water molecule in its own liquid of $-7.09\pm0.03$~kcal\,mol$^{-1}$, consistent
to $\lesssim\!0.19$~kcal\,mol$^{-1}$ with three independent literature values: regularized
potential-distribution theory~\cite{WeberMerchantAsthagiri2011} ($-7.0$), quasi-chemical
theory~\cite{MerchantAsthagiri2011} ($-6.9$), and Bennett-acceptance-ratio FEP~\cite{ShirtsPande2005}
($-7.05$, the closest). Assembled into a
standard molar entropy at the liquid density via $S^\circ = S^{\rm ideal}_{\rm tr+rot} +
(U_{\rm inter}-\mu_{\rm ex})/T - R$, the trailing $-R$ being the residual-pressure term that converts
the excess chemical potential to the excess Helmholtz energy\footnote{At liquid density the residual
pressure $P-\rho k_BT\approx-\rho k_BT$, so $A_{\rm ex}=\mu_{\rm ex}+RT$ and the excess (residual)
entropy is $S_{\rm ex}=(U_{\rm inter}-\mu_{\rm ex})/T-R$; omitting the $-R$ overestimates
$S^\circ_{\rm FEP}$ by exactly $R=8.31$~J\,mol$^{-1}$K$^{-1}$.}, this yields $S^\circ_{\rm FEP}=65.1\pm0.4$~J\,mol$^{-1}$K$^{-1}$
(the three literature $\mu_{\rm ex}$ give $62$--$65$); the 3PT value ($64.0$) sits
$1.1$~J\,mol$^{-1}$K$^{-1}$ below this own-FEP anchor, a larger residual than the $0.1$ found for
TIP4P/2005. This remaining translational deficit is analyzed in Sec.~\ref{sec:lawnm}. As another point of comparison, we note that before the Yeh--Hummer correction, the 3PT value is $63.5$, close to the FEP-derived literature entropy of SPC/E, $63.36$, assembled
by Lin \emph{et al.}\cite{Lin2010} from the Shirts--Pande free energies\cite{ShirtsPande2005}. That the same trans+rot cage, with no adjustable parameter, transfers across two independent
water models, matching the rigorous free-energy entropy for TIP4P/2005 and capturing most of it for
SPC/E, is evidence that the construction captures real hydration-shell physics rather than the
conventions of a single force field.

\begin{table}[t]
\caption{\label{tab:waterB}Standard molar entropy of liquid SPC/E water at 298~K
(J\,mol$^{-1}$K$^{-1}$), from one consistent 1728-molecule computation; the translational and
rotational (libration) cages bring the rigorous-HS baseline onto the independent FEP self-solvation
value.}
\begin{ruledtabular}
\begin{tabular}{lcl}
Method & $S^\circ_{\rm liq}$ & note \\
\colrule
rigorous-HS 2PT (uncorrected) & 57.1 & baseline, no correction \\
3PT ($+$ translational cage) & 58.5 & $\Scage^{\rm tr}=+1.4$ \\
canonical-2PT & 60.8 & empirical $+\ln Z$ \\
3PT (translational $+$ rotational cage) & \textbf{64.0} & trans $+1.4$, rot $+5.0$ ($+0.5$ Yeh--Hummer) \\
FEP self-solvation ($\mu_{\rm ex}=-7.09\pm0.03$) & $65.1\pm0.4$ & this work (five-seed BAR); lit.\ $-6.9$ to $-7.05$ give $62$--$65$ \\
experiment (CODATA) & 69.95 & classical$+$quantum \\
\end{tabular}
\end{ruledtabular}
\end{table}

%==============================================================
\section{Discussions}\label{sec:discussions}
%==============================================================

\subsection{The cage is not a harmonic solid}\label{sec:physics}
\begin{figure}[t]
\includegraphics[width=\columnwidth]{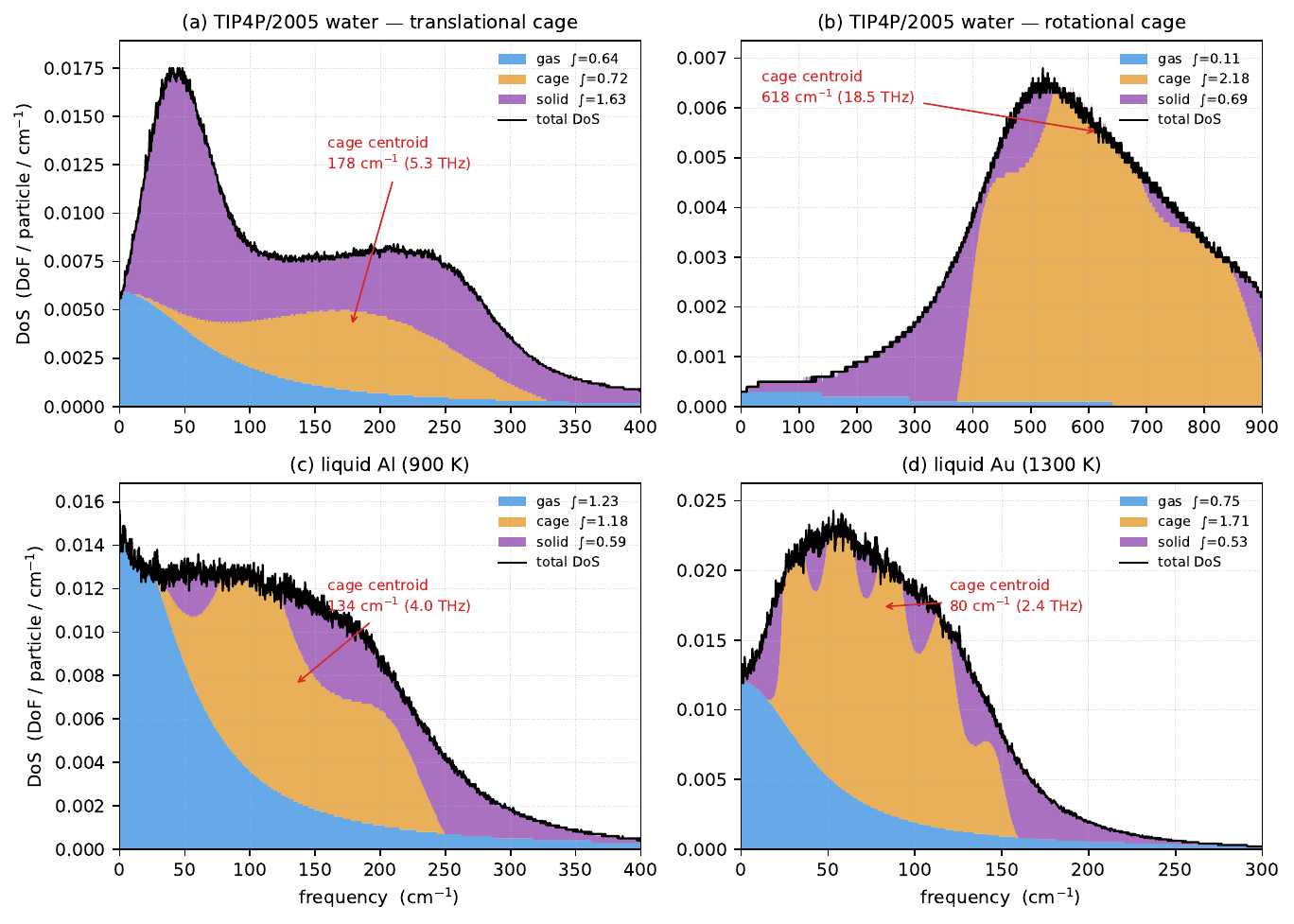}
\caption{\label{fig:cagedos}Gas$|$cage$|$solid decomposition of the velocity DoS (stacked, summing to
the total) for four liquids: (a) TIP4P/2005 water, translational; (b) water, rotational; (c) liquid
Al ($900$~K); (d) liquid Au ($1300$~K). The cage (orange) is the non-Markovian memory band carved
from the solid. Water's rotational channel is cage-dominated ($\int$cage$=2.18$ of $3$, centroid
$\sim618$~cm$^{-1}$); its translational cage is a distinct higher-frequency rattle ($\int=0.72$, centroid
$\sim178$~cm$^{-1}$) lying above the harmonic H-bond solid ($\sim52$~cm$^{-1}$). Metal cages span
$\sim80$~cm$^{-1}$ (soft, heavy Au) to $\sim134$~cm$^{-1}$ (Al). Band positions are reported as
cage-band centroids (first moments), which are robust where the broad cage maximum is not. $x$-axis
in cm$^{-1}$; THz equivalents ($1$~THz${=}33.36$~cm$^{-1}$) in the panel annotations.}
\end{figure}

3PT naturally affords a decomposition of the velocity DoS into gas, cage
and harmonic-solid channels (Fig.~\ref{fig:cagedos}): the non-Markovian cage appears as a distinct
band, carved from the solid and separate from the diffusive gas ($\nu\!\to\!0$). Liquid metals carry a large translational cage: $1.18$ of $3$ DoF in Al (centroid $\sim134$~cm$^{-1}$),
rising to $1.71$ in soft, heavy Au (centroid $\sim80$~cm$^{-1}$, cage-dominated). Here, the cage frequency
tracks $\sqrt{k/m}$ (heavy, soft Au lowest; light Al highest), consistent with a real,
mass-dependent cage band across the metals series rather than a water-specific artifact. 

In liquid water it
is the rotational channel that is more cage-dominated: the cage carries $2.18$ of the $3$
rotational degrees of freedom, in the hindered-libration band centered near $\sim618$~cm$^{-1}$
($18.5$~THz), against only $0.69$ truly harmonic. Therefore, most of what would be classified as rotational
``solid-like'' motion in 2PT is in fact caged free-rotor rattling, and it supplies the
larger molecular entropy increment, $+5.2$ versus $+1.4$~J\,mol$^{-1}$K$^{-1}$. This is the expected
signature of a deep orientational H-bond cage: a nearly free rotor would place its weight near
$\nu\!\to\!0$, whereas water fills the libration band instead. The translational
cage is a distinct higher-frequency rattle ($0.72$ DoF, band centered near $\sim178$~cm$^{-1}$), lying above the harmonic H-bond solid, which peaks near $\sim52$~cm$^{-1}$. The broad, flat-topped band makes the bare maximum noise-sensitive, so we report the robust centroid.

The above peak assignment for water may appear at first to be counterintuitive: it places the higher-frequency band in the anharmonic cage and the lower in
the harmonic solid. This is the reverse of a frequency-stiffness reading, in which the soft low-frequency
mode would seem the anharmonic one. The key reconciliation is to note that the cage is defined by memory, not stiffness. Indeed, it is the
non-Markovian backscattering excess (the negative VACF dip), the motion of a molecule rattling
directly against its hydrogen-bonded neighbor. This motion is consistent with the frequency of the H-bond stretch. The
slower O$\cdots$O$\cdots$O bend is plausibly the transverse, restoring-force network mode that the
construction returns to the harmonic solid. In this light, our band assignment aligns with measurements: liquid water has two well-characterized intermolecular
translational bands, the O$\cdots$O$\cdots$O bend near $\sim60$~cm$^{-1}$ and the O$\cdots$O
hydrogen-bond stretch near $\sim180$--$200$~cm$^{-1}$ (far-IR, Raman and inelastic neutron
scattering~\cite{Persson2017}). The 3PT cage, knowing only the VACF, places its band
(centroid $\sim178$~cm$^{-1}$) on the measured H-bond stretch and the residual solid band
(peak $\sim52$~cm$^{-1}$) on the bend. The agreement of band positions is further validation of the partition; however, a rigorous validation of mode character (e.g., projecting onto the inelastic-neutron dynamic structure factor or a normal-mode participation analysis) is beyond our scope and left to future work. 

\subsection{Cage efficiency arises from kernel non-Markovianity}\label{sec:lawnm}
We argue that the LJ under-prediction, the metals' exact match, and the alignment with formal free-energy calculations in water can be unified through a single scaling curve. We define the \emph{cage efficiency} as the fraction of the rigorous-HS
deficit that the explicit cage recovers:
\begin{equation}
  \eta_{\rm cage} \equiv \frac{\Delta S_{\rm cage}}{S_{\rm ref}-S_{\rm rigHS}},
  \label{eq:eff}
\end{equation}
and we correlate this efficiency against the \emph{non-Markovianity} of the kernel, a dynamical order parameter defined as $\gamma/\Omega_0$, the ratio of the zero-frequency
friction $\gamma=\tilde K(0)$ to the Einstein frequency $\Omega_0$ (i.e., the second moment of the total
DoS). This order parameter is a dimensionless damping parameter of the generalized Langevin equation: $\gamma/\Omega_0\!\ll\!1$
is an underdamped, nearly ballistic (Markovian) rattle, while $\gamma/\Omega_0\!\gg\!1$ is an
overdamped, strongly memory-laden cage. Critically, $\gamma/\Omega_0$ is built from the bare kernel
and DoS alone. It uses none of the cage clip, gate, filter, or weight $(W_g-W_s)$, so its
correlation with $\eta_{\rm cage}$ is not circular.

\begin{figure}[tbp]
\includegraphics[width=\columnwidth]{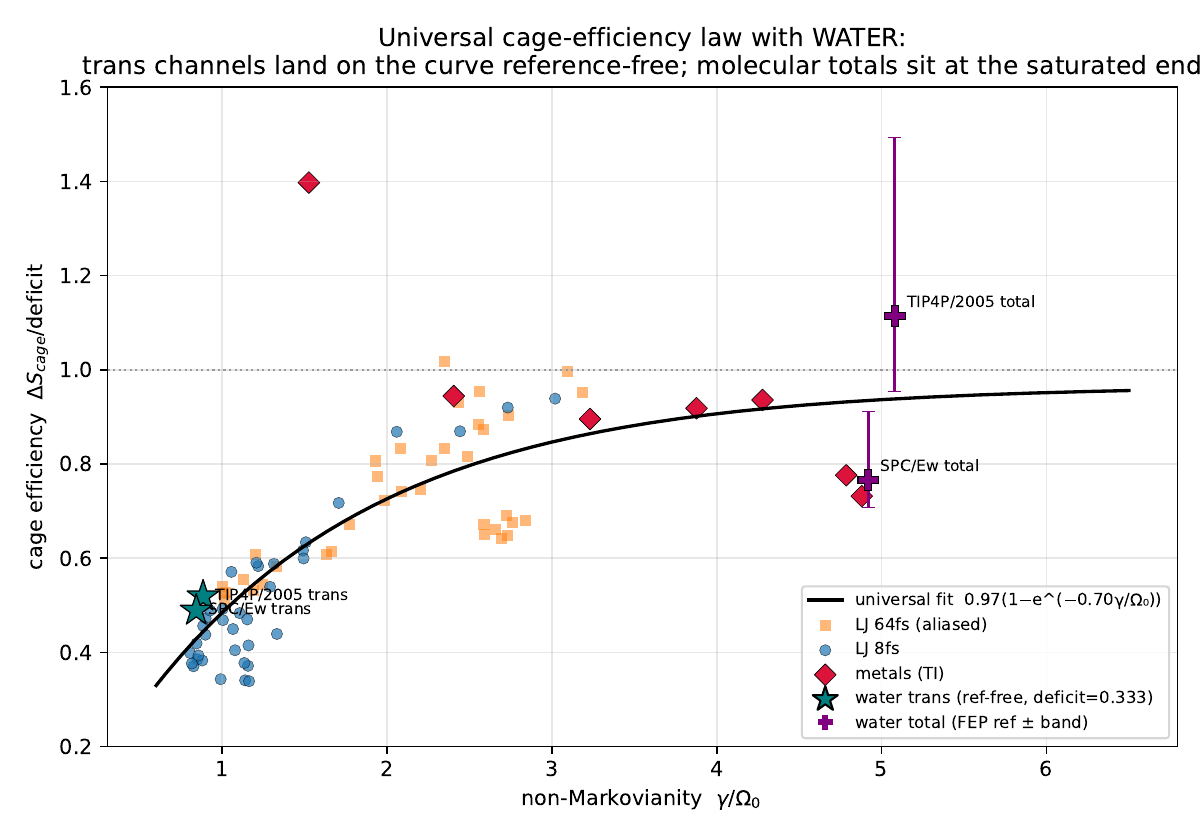}
\caption{\label{fig:lawnm}Cage efficiency $\eta_{\rm cage}$ (Eq.~\eqref{eq:eff}) versus kernel
non-Markovianity $\gamma/\Omega_0$, showing an empirical scaling curve (Eq.~\eqref{eq:law}; black) across LJ
liquids (blue, $8$~fs), the seven Sutton--Chen metals (red, vs TI), and water (translational
channels as stars, placed reference-free with the universal $0.333\,\kB$ deficit; molecular totals
as crosses with the free-energy-reference band). Orange squares show the same LJ states sampled at
$64$~fs: coarse sampling fabricates kernel memory and slides each state \emph{up} the same curve.
Metals and the water rotational channel sit on or near the saturated plateau
($\eta_{\rm cage}\!\to\!0.97$); weakly non-Markovian LJ liquids and the water translational
channels populate the rising flank.}
\end{figure}

Two facts make $\eta_{\rm cage}$ informative. First, the deficit in the denominator of
Eq.~\eqref{eq:eff} is nearly a constant of the material: across the LJ liquid grid and the seven
metals it is $0.333\pm0.036\,\kB$ per particle and is uncorrelated with $\gamma/\Omega_0$
(Pearson $+0.05$). The missing entropy is a near-universal anharmonic increment; what varies is how
much of it is carried as kernel memory. Its magnitude, $\approx\!1\,\kB$/atom in three dimensions,
matches the scale of the near-universal liquid entropy constant identified by Wallace for monatomic
liquids\cite{Wallace1997,Wallace2002}, suggesting that the constancy of the deficit is not an
artifact of our baseline but the same universal anharmonic entropy, here measured from the
rigorous-hard-sphere reference. Second, that memory fraction, and hence
$\eta_{\rm cage}$, follows a saturating scaling curve in $\gamma/\Omega_0$ for LJ and metals
(Fig.~\ref{fig:lawnm}); a convenient summary is
\begin{equation}
  \eta_{\rm cage}\;\approx\;0.97\,\bigl(1-e^{-0.70\,\gamma/\Omega_0}\bigr),
  \label{eq:law}
\end{equation}
with $R^2=0.63$ over the pooled $81$ states (the $37$ LJ liquids for which resolution-converged
8-fs kernels were computed, the same states re-sampled at 64~fs, and the seven metals). The fit is insensitive to this pooling: restricting it
to the resolution-converged 8-fs LJ and metal states alone gives
$0.98\,(1-e^{-0.66\,\gamma/\Omega_0})$ ($R^2=0.60$), and a bootstrap over states brackets the
plateau at $[0.88,1.27]$ and the rate at $[0.44,0.82]$ ($95\%$). The residual scatter reflects
state-to-state noise in the small cage rather than a departure from the trend. 

We emphasize that
the exponential form of Eq.~\eqref{eq:law} is a convenient interpolation, not a derived law; its constrained content is
the two limits it connects. At weak memory the cage vanishes superlinearly: the Einstein-frequency
filter suppresses the high-frequency excess of short-memory kernels, so weakly non-Markovian
systems draw very little cage. Where the deficit is fully memory-borne, the efficiency
saturates near unity, with the plateau level anchored empirically by the metals and the water
rotational channel. The curve should therefore be read as a property of liquid-state (escape-coupled)
kernels rather than of the construction itself: kernel memory without low-frequency structural
relaxation does not produce a cage. The plateau near unity indicates that where the deficit
is heavily memory-borne ($\gamma/\Omega_0\!\gtrsim\!3$), the cage recovers essentially all of
it. The rising flank ($\gamma/\Omega_0\!\sim\!1$) represents systems where the deficit stems from soft anharmonicities rather than memory; here, the cage correctly supplies only that memory-borne fraction. Moreover, the primary, parameterization-independent evidence is a weight-free memory metric: the relative spectral area $\int|\FK-\FM|/\int\FK$, built
from the bare kernel alone, predicts $\eta_{\rm cage}$ more tightly than $\gamma/\Omega_0$ itself
($r=0.79$), confirming that it is genuinely the kernel's memory content, not a coincidence of any
single summary variable, that governs the cage.

The scaling curve unifies the three liquid material classes. Sutton--Chen metals (characterized by many-body
electronic caging; $\gamma/\Omega_0\!\approx\!1.5$--$4.9$) and the water rotational channel (the
rigid hydrogen-bond network; $\gamma/\Omega_0\!\approx\!6$) are strongly non-Markovian and sit on
or near the saturated plateau, allowing 3PT to reproduce their TI and FEP reference entropies nearly exactly. The
water translational channel is only moderately non-Markovian
($\gamma/\Omega_0\!\approx\!0.85$ once the main-lobe friction safeguard of
Sec.~\ref{SI-app:scope} is applied): it lies on the rising flank, where its partial cage recovery
falls on Eq.~\eqref{eq:law} reference-free, using only the universal $0.333\,\kB$ deficit. This flank placement is the same memory shortfall behind the translational
residual of Sec.~\ref{sec:water}. Monoatomic LJ liquids at moderate density are only
weakly non-Markovian ($\gamma/\Omega_0\!\approx\!1$ at $\rho^*\!=\!0.72$, rising to $\sim\!2.7$ near
freezing), so they populate the rising flank. The LJ under-correction of Sec.~\ref{sec:lj} is
exactly $\eta_{\rm cage}<1$ read off that flank, climbing from $0.34$ to $0.92$ with density. The
ratio $\gamma/\Omega_0$ thus serves as an \emph{a priori} diagnostic for the quantitative domain of the 3PT functional.

The scaling curve also resolves a sampling subtlety. The kernel's non-Markovianity is read from the fast
initial decay of the velocity autocorrelation function; sampling the trajectory too coarsely
under-resolves that decay and makes the inferred kernel appear smoother and longer-correlated,
artificially inflating $\gamma/\Omega_0$ and the cage. On the same LJ states, $64$-fs sampling
displaces every point along the universal curve toward higher $\gamma/\Omega_0$ and higher
$\eta_{\rm cage}$; refining to $8$~fs slides $35$ of $37$ states back down the same curve
(Fig.~\ref{fig:lawnm}). The $8$-fs values used throughout are the resolution-converged ones (a
control at $2$~ns confirms the difference is the timestep, not the trajectory length).

\subsection{Trans-dimensional evidence for the prefactor}\label{sec:universality}
\begin{figure}[tbp]
\includegraphics[width=\columnwidth]{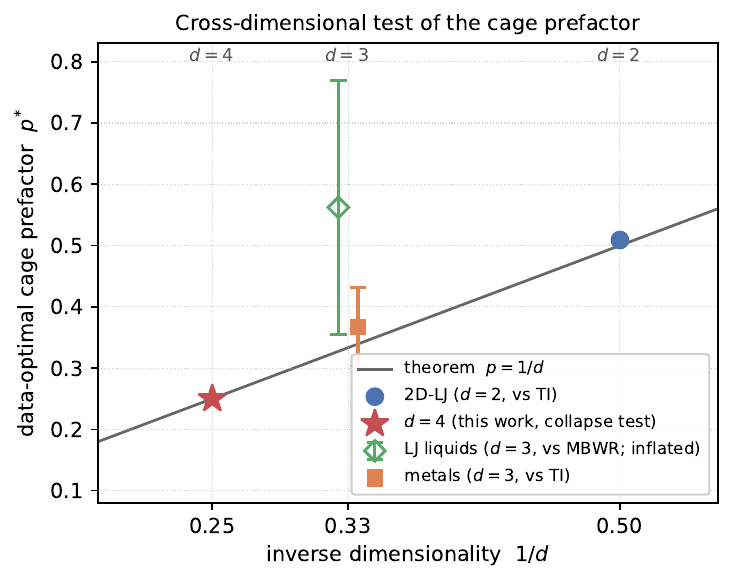}
\caption{\label{fig:collapse}Data-optimal cage prefactor $p^*$ versus inverse dimensionality $1/d$,
against the predicted line $p=1/d$. The clean readings are the strongly-caged metals
($0.37\pm0.06$, vs TI) and the 2D-LJ liquid ($p^*=0.51$); the 3D LJ liquids give an inflated
$0.56\pm0.21$ (open symbol) because their weak non-Markovianity forces $p^*$ to absorb the
cage's memory shortfall (Sec.~\ref{sec:lawnm}), so they do not constitute a clean dimensional
reading. The $d=4$ value $1/4$, predicted by the same argument, is supported by a
four-dimensional LJ liquid (Sec.~\ref{sec:protocol}; SI Fig.~\ref{SI-fig:d4prefactor}).}
\end{figure}

The data-optimal prefactor $p^*$, the single constant that centers $S_{\rm 3PT}$ on the
reference equation of state, is a clean probe of dimensionality only where the cage is on its
saturated plateau. Thus, $p^*$ reflects $d$ rather than missing memory. For the seven
Sutton--Chen metals (strongly non-Markovian, vs TI) it is $p^*=0.37\pm0.06$, straddling $1/3$;
for a dense two-dimensional LJ liquid ($\rho^*=0.85$, $T^*=1.0$, vs 2D-LJ thermodynamic
integration) it is $p^*=0.51\approx1/2$ (Fig.~\ref{fig:collapse}). The monoatomic-LJ liquids, on the other hand, do not provide a clean reading. Their $p^*$ is inflated to $0.56\pm0.21$ and trends with density, rising from $\sim\!0.3$ near freezing to $>\!0.6$ at moderate density. On this weakly non-Markovian flank (Sec.~\ref{sec:lawnm} above), the data-optimal prefactor is forced to absorb the portion of the nearly density-independent deficit that the fixed $p=1/3$ cage cannot capture. It conflates dimensionality with the memory shortfall and is therefore not a reliable dimensional probe. Thus the LJ point in
Fig.~\ref{fig:collapse} is shown for completeness with this caveat. That $d=2$ requires
$p\!\approx\!1/2$ and not $1/3$ is instead shown decisively by a residual test independent of any
fit: applying $p=1/2$ lands the 2D-LJ entropy on its TI reference to $0.008\,\kB$, whereas
$p=1/3$ misses by $0.15\,\kB$ (a $\sim19\times$ larger residual). 

The cross-dimensional sequence closes with the predicted $d=4$ value $p=1/4$, which we tested directly on a
genuinely 4D LJ liquid (Sec.~\ref{sec:protocol}). However, here $d=4$ does not even afford the
clean residual test of $d=2$: the soft-mode anharmonic deficit scales with the excess entropy
and is several times larger at $d=4$. This contaminates the data-optimal $p^*$ even on the caged plateau,
inflating the bare $S_{\rm 3PT}$-centering value to $\sim\!1.3$, even further from $1/4$ than the inflated $0.56$ of the $d=3$ liquids is from $1/3$. We therefore use the efficiency
collapse itself as the discriminator: a candidate prefactor is admissible only if the residual non-memory
entropy, i.e., the part of the deficit the cage does not carry once $\Delta S_{\rm
cage}(p)$ is placed on the universal $\eta(\mathcal{M})$ curve, is positive and flat across
state points. This is a good figure of merit, as the expected non-memory leftover is a fixed per-mode anharmonic under-correction (the same memory-independent residual the
cage leaves on the $d=3$ liquids). We find that over the $22$ 4D fluid
states (the $T^*=3.0$ and $4.0$ isotherms and the $P^*=3.2$ isobar, memory area
$\mathcal{M}=0.17$--$0.72$), $p=1/4$ leaves a residual that is flat at $0.10\,\kB$/mode (coefficient of
variation $14\%$), while $p=1/2$ both collapses to $\approx0.04\,\kB$/mode and acquires a three-fold larger,
systematically $\mathcal{M}$-dependent scatter (it over-subtracts at the strongly caged states), and can be
excluded. We find that $p=1/3$ is intermediate, with a residual of $0.08\,\kB$/mode, but mildly sloped (SI Fig.~\ref{SI-fig:d4prefactor}).
The flat $\sim\!0.1\,\kB$/mode that $p=1/4$ leaves is the per-mode counterpart of the
density-independent anharmonic under-correction of Sec.~\ref{sec:lawnm}, magnified at $d=4$ in
proportion to the larger excess entropy. The 4D liquid thus selects $p=1/4$ over $p=1/2$
and favors it over $p=1/3$, completing the sequence $p=1/d$ for $d=2,3,4$. We note that this is a consistency-level
test, since the universal $\eta(\mathcal{M})$ curve is calibrated on lower-dimensional data. An
explicit 4D model of the anharmonic residual would sharpen the argument further, and is left for future work.

\subsection{3PT yields a smooth, empirically integrable entropy surface}\label{sec:statefn}
\begin{figure}[tbp]
\includegraphics[width=\columnwidth]{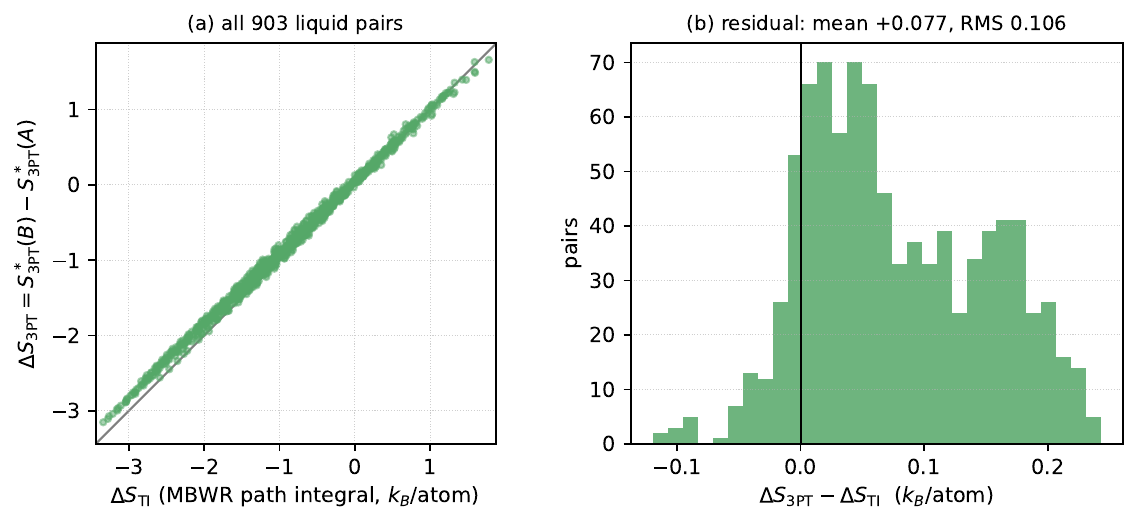}
\caption{\label{fig:statefn}3PT as a state function. (a)~Entropy difference $\Delta S_{\rm 3PT}$
versus the MBWR thermodynamic-integration $\Delta S_{\rm TI}$ for all 903 pairs of the 43 liquid
states (line $y=x$). (b)~Residual $\Delta S_{\rm 3PT}-\Delta S_{\rm TI}$ (mean $+0.077$, RMS
$0.106\,\kB$); the RMS is well below $\sqrt2\times$ the single-point error, so point deviations
largely cancel in differences, indicating a smooth surface with a coherent density-dependent offset.}
\end{figure}

Because 3PT assigns an absolute entropy to each state point from a single trajectory with no
reference state, the independent point values must lie on a single integrable surface if the method
is thermodynamically sound. We verify this for the LJ liquids dataset three ways. (i)~The entropy difference
$\Delta S_{\rm 3PT}=S^*_{\rm 3PT}(B)-S^*_{\rm 3PT}(A)$ between every pair of the 43 liquid states
reproduces the thermodynamic-integration difference $\Delta S_{\rm TI}$ from the MBWR equation of
state (Fig.~\ref{fig:statefn}a): over all 903 pairs the residual is $0.106\,\kB$ RMS (mean $+0.077$),
even for $\Delta S$ spanning $\sim\!5\,\kB$ across distant states. This is well below
$\sqrt2\times$ the single-point RMS ($0.16\,\kB$), so the point errors largely cancel in the
difference, which, in turn, is the signature of a smooth surface with a systematic offset rather than random scatter
(Fig.~\ref{fig:statefn}b). (ii)~The reference is itself a consistent surface: integrating its
$(\partial S/\partial\rho)_T$ and $(\partial S/\partial T)_\rho=C_v/T$ along two different paths
(isotherm-then-isochore vs the reverse) agrees to $\sim\!10^{-9}\,\kB$, so $\Delta S_{\rm TI}$ is
path-independent, as any state function must be. (iii)~The 43 independent 3PT values are fit by a
single quadratic $S(\rho^*,T^*)$ surface with residual $0.025\,\kB$, of the same order as the MBWR
EOS itself ($0.011\,\kB$), showing that the independently estimated 3PT entropies lie on a smooth and low-residual empirical surface over the LJ liquid grid, rather than reflecting only local agreement.
The $+0.077\,\kB$ mean tilt in the residual is the coherent, density-dependent under-correction of
Sec.~\ref{sec:lawnm}, a smooth bias of the memory-limited cage rather than a loss of integrability. The per-point deviation trends from $-0.24$ at $\rho^*=0.70$ to $\approx-0.06$
near $\rho^*=1.0$.

\FloatBarrier
%==============================================================
\section{Conclusions}\label{sec:conclusions}
%==============================================================
In summary, we have introduced 3PT as a memory-resolved extension of
density-of-states thermodynamics. The central result is to separate the finite-memory
part of liquid motion from the harmonic remainder: the cage is obtained directly
from the Mori--Zwanzig kernel as the non-Markovian excess of the full velocity
response over its Markovian reference. Assigning this trajectory-defined spectrum
an excluded-volume entropy reference, with the dimensional projection factor
$p=1/d$ and a fluidicity gate that restores the crystalline limit, yields a
gas--cage--solid decomposition that preserves the usual limiting behavior while
making the anharmonic liquid contribution explicit.

Taken together, the benchmarks show that the cage is not a system-specific
correction but a transferable dynamical component of the entropy. Systems with
strong kernel memory, including liquid metals and hydrogen-bonded water, are
described quantitatively without fitting to their reference entropies, whereas
weakly non-Markovian Lennard-Jones liquids are corrected only partially. This
separation is useful: the $\gamma/\Omega_0$ order parameter becomes an a priori diagnostic
of whether the missing entropy is expected to reside primarily in the memory cage or
in softer, non-memory anharmonic modes. Thus 3PT does not merely improve a
numerical estimator; it identifies which part of the liquid entropy is dynamically
encoded in the memory kernel.

The present formulation should be most reliable for equilibrium liquids in which
diffusive and cage motions remain spectrally separable. Regimes where those
processes merge, such as deeply supercooled liquids, may require a simultaneous
rather than sequential gas--cage decomposition. Likewise, confined or strongly
anisotropic fluids provide a natural test of the dimensional prefactor, since
broken isotropy should move the cage weighting away from the bulk value
$p=1/d$. More broadly, the results suggest that liquid entropy can be organized
around dynamical memory: the harmonic solid is not the default reservoir for all
non-diffusive spectral weight, and the transient cage is a distinct
thermodynamic component that can be read directly from an equilibrium trajectory.

\begin{acknowledgments}
This work was supported by the US National Science Foundation through grant CBET-2145144. This work used
resources of the National Energy Research Scientific Computing Center, a DOE Office of Science User
Facility supported by the Office of Science of the U.S. Department of Energy under Contract
No.\ DE-AC02-05CH11231 and the m3047 allocation on the Perlmutter machine. This work used the
Extreme Science and Engineering Discovery Environment (XSEDE) and the Expanse supercomputer at the
San Diego Supercomputer Center, supported by National Science Foundation grant number ACI-1548562
under the DDP381 allocation. T.A.P.\ acknowledges support from the Alfred P.\ Sloan Foundation.
\end{acknowledgments}

\section*{Supporting Information}
See the supporting information for: the full derivation of the cage prefactor $p=1/d$ (Sec.~\ref{SI-app:derivation}); the hard-sphere entropy reference and the scope of the gas--cage extraction (Sec.~\ref{SI-app:scope}); the spectral filter, cutoff sensitivity, and fluidicity gate (Sec.~\ref{SI-app:filter}); the rotational cage entropy derivation (Sec.~\ref{SI-app:rot3pt}); the explicit four-site free-energy-perturbation determination of the water self-solvation free energy (Sec.~\ref{SI-app:fep}); the Frenkel--Ladd/Einstein-molecule absolute entropy of TIP4P/2005 ice (Sec.~\ref{SI-app:ti}); and the complete Lennard-Jones $(\rho^*,T^*)$ grid of entropies (Sec.~\ref{SI-app:grid}, Table~\ref{SI-tab:s1}).

\section*{Data Availability Statement}
The data that support the findings of this study are available at our group's online repository (\url{https://github.com/atlas-nano/manuscript-data/tree/main/3pt-construction}) and are archived on Zenodo at \url{https://doi.org/10.5281/zenodo.21447740}. The Python-based code (py-xPT) that implements 2PT/3PT and used for all our analysis is available at \url{https://github.com/atlas-nano/codes/tree/main/py-xPT} and archived at \url{https://doi.org/10.5281/zenodo.21447746}. The 4D Lennard-Jones simulator is available at \url{https://github.com/atlas-nano/codes/tree/main/lj-4d-md} and archived at \url{https://doi.org/10.5281/zenodo.21447750}.

\bibliography{refs}

@article{Plimpton1995,
  title = {Fast parallel algorithms for short-range molecular dynamics},
  author = {Plimpton,  Steve},
  journal = {J. Comput. Phys.},
  volume = {117},
  number = {1},
  pages = {1--19},
  year = {1993},
  doi = {10.1006/jcph.1995.1039},
}

@article{Lin2003,
  title = {The two-phase model for calculating thermodynamic properties of liquids from molecular dynamics: Validation for the phase diagram of Lennard-Jones fluids},
  author = {Lin, Shiang-Tai and Blanco, Mario and Goddard, III, William A},
  journal = {The Journal of Chemical Physics},
  volume = {119},
  number = {22},
  pages = {11792--11805},
  year = {2003},
  doi = {10.1063/1.1624057},
  publisher = {AIP Publishing},
}

@article{Lin2010,
  title = {Two-phase thermodynamic model for efficient and accurate absolute entropy of water from molecular dynamics simulations},
  author = {Lin, Shiang-Tai and Maiti, Prabal K and Goddard, 3rd, William A},
  journal = {Journal of Physical Chemistry B},
  volume = {114},
  number = {24},
  pages = {8191--8198},
  year = {2010},
  doi = {10.1021/jp103120q},
  pmid = {20504009},
  publisher = {American Chemical Society (ACS)},
}

@article{Pascal2011,
  title = {Thermodynamics of liquids: standard molar entropies and heat capacities of common solvents from 2PT molecular dynamics},
  author = {Pascal, Tod A and Lin, Shiang-Tai and Goddard, 3rd, William A},
  journal = {Physical Chemistry, Chemical Physics - PCCP},
  volume = {13},
  number = {1},
  pages = {169--181},
  year = {2011},
  doi = {10.1039/c0cp01549k},
  pmid = {21103600},
  publisher = {Royal Society of Chemistry (RSC)},
}

@article{Desjarlais2013,
  title = {First-principles calculation of entropy for liquid metals},
  author = {Desjarlais,  Michael P.},
  journal = {Physical review. E, Statistical, nonlinear, and soft matter physics},
  volume = {88},
  number = {6},
  pages = {062145},
  year = {2013},
  doi = {10.1103/PhysRevE.88.062145},
  pmid = {24483423},
  publisher = {American Physical Society (APS)},
}

@article{Sun2017,
  title = {Two-phase thermodynamic model for computing entropies of liquids reanalyzed},
  author = {Sun,  Tao and Xian,  Jiawei and Zhang,  Huai and Zhang,  Zhigang and Zhang,  Yigang},
  journal = {Journal of Chemical Physics},
  volume = {147},
  number = {19},
  pages = {194505},
  year = {2017},
  doi = {10.1063/1.5001798},
  pmid = {29166119},
  publisher = {AIP Publishing},
}

@article{Johnson1993,
  title = {The Lennard-Jones equation of state revisited},
  author = {Johnson,  J. Karl and Zollweg,  John A. and Gubbins,  Keith E.},
  journal = {Molecular Physics},
  volume = {78},
  number = {3},
  pages = {591--618},
  year = {1993},
  doi = {10.1080/00268979300100411},
  publisher = {Informa UK Limited},
}

@article{vdHoef2000,
  title = {Free energy of the Lennard-Jones solid},
  author = {Hoef, M. A.},
  journal = {J. Chem. Phys.},
  volume = {113},
  number = {18},
  pages = {8142--8148},
  year = {2000},
  doi = {10.1063/1.1314342},
}

@article{AbascalVega2005,
  title = {A general purpose model for the condensed phases of water: TIP4P/2005},
  author = {Abascal, J L F and Vega, C},
  journal = {The Journal of Chemical Physics},
  volume = {123},
  number = {23},
  pages = {234505},
  year = {2005},
  doi = {10.1063/1.2121687},
  pmid = {16392929},
  publisher = {AIP Publishing},
}

@article{Berendsen1987,
  title = {The missing term in effective pair potentials},
  author = {Berendsen,  H. J. C. and Grigera,  J. R. and Straatsma,  T. P.},
  journal = {The Journal of Physical Chemistry},
  volume = {91},
  number = {24},
  pages = {6269--6271},
  year = {1987},
  doi = {10.1021/j100308a038},
  publisher = {American Chemical Society (ACS)},
}

@article{ShirtsPande2005,
  title = {Solvation free energies of amino acid side chain analogs for common molecular mechanics water models},
  author = {Shirts,  Michael R. and Pande,  Vijay S.},
  journal = {Journal of Chemical Physics},
  volume = {122},
  number = {13},
  pages = {134508},
  year = {2005},
  doi = {10.1063/1.1877132},
  pmid = {15847482},
  publisher = {AIP Publishing},
}

@article{WeberMerchantAsthagiri2011,
  title = {Communication: Regularizing binding energy distributions and thermodynamics of hydration: Theory and application to water modeled with classical and ab initio simulations},
  author = {Weber,  Val'{e}ry and Merchant,  Safir and Asthagiri,  D.},
  journal = {Journal of Chemical Physics},
  volume = {135},
  number = {18},
  pages = {181101},
  year = {2011},
  doi = {10.1063/1.3660205},
  pmid = {22088043},
  publisher = {AIP Publishing},
}

@article{MerchantAsthagiri2011,
  title = {Water coordination structures and the excess free energy of the liquid},
  author = {Merchant,  Safir and Shah,  Jindal K. and Asthagiri,  D.},
  journal = {Journal of Chemical Physics},
  volume = {134},
  number = {12},
  pages = {124514},
  year = {2011},
  doi = {10.1063/1.3572058},
  pmid = {21456683},
  publisher = {AIP Publishing},
}

@article{Mori1965,
  title = {Transport, collective motion, and Brownian motion},
  author = {Mori, Hazime},
  journal = {Progress of theoretical physics},
  volume = {33},
  number = {3},
  pages = {423--455},
  year = {1965},
  publisher = {Oxford University Press (OUP)},
  doi = {10.1143/PTP.33.423},
}

@book{frenkel_smit_2002,
  author = {Frenkel, Daan and Smit, Berend},
  title = {Understanding Molecular Simulation: From Algorithms to Applications},
  edition = {2nd},
  publisher = {Academic Press},
  year = {1996},
  doi = {10.1063/1.881812},
}

@misc{py-xPT_2026_commit,
  author = {{atlas-nano}},
  title = {{py-xPT: Python implementation of 1PT/2PT/3PT Methods (repository)}},
  howpublished = {\url{https://github.com/atlas-nano/codes/tree/main/py-xPT}},
  year = {2026},
  doi = {10.5281/zenodo.21447746},
  note = {Commit a400326bbfdb5ee608337d3cd7a16bbcbc9a8141 (release tag 3pt-v1); archived at Zenodo, \url{https://doi.org/10.5281/zenodo.21447746}; accessed 2026-07-19},
}

@article{Mohr2005,
  title = {CODATA recommended values of the fundamental physical constants: 2002},
  author = {Mohr, Peter J and Taylor, Barry N},
  journal = {Reviews of modern physics},
  volume = {77},
  number = {1},
  pages = {1--107},
  year = {2005},
  publisher = {American Physical Society (APS)},
  doi = {10.1103/revmodphys.77.1},
}

@article{Parrinello1981,
  title = {Polymorphic transitions in single crystals: A new molecular dynamics method},
  author = {Parrinello, Michele and Rahman, Aneesur},
  journal = {Journal of Applied physics},
  volume = {52},
  number = {12},
  pages = {7182--7190},
  year = {1981},
  publisher = {AIP Publishing},
  doi = {10.1063/1.328693},
}

@article{Martyna1994,
  title = {Constant pressure molecular dynamics algorithms},
  author = {Martyna, Glenn J and Tobias, Douglas J and Klein, Michael L},
  journal = {The Journal of Chemical Physics},
  volume = {101},
  number = {5},
  pages = {4177-4189},
  year = {1994},
  doi = {10.1063/1.467468},
  publisher = {AIP Publishing},
}

@article{Shinoda2004,
  title = {Rapid estimation of elastic constants by molecular dynamics simulation under constant stress},
  author = {Shinoda, Wataru and Shiga, Motoyuki and Mikami, Masuhiro},
  journal = {Physical Review B},
  volume = {69},
  number = {13},
  pages = {134103},
  year = {2004},
  doi = {10.1103/PHYSREVB.69.134103},
  publisher = {American Physical Society (APS)},
}

@article{Sutton1990,
  title = {Long-range finnis--sinclair potentials},
  author = {Sutton, AP and Chen, J},
  journal = {Philosophical Magazine Letters},
  volume = {61},
  number = {3},
  pages = {139--146},
  year = {1990},
  publisher = {Informa UK Limited},
  doi = {10.1080/09500839008206493},
}

@article{Tuckerman2006,
  title = {A Liouville-operator derived measure-preserving integrator for molecular dynamics simulations in the isothermal--isobaric ensemble},
  author = {Tuckerman, M. and Alejandre, J. and López-Rendón, R. and Jochim, Andrea L. and Martyna, G.},
  journal = {Journal of Physics A: Mathematical and General},
  volume = {39},
  number = {19},
  pages = {5629--5651},
  year = {2006},
  doi = {10.1088/0305-4470/39/19/S18},
  publisher = {IOP Publishing},
}

@article{Thompson2022,
  title = {LAMMPS-a flexible simulation tool for particle-based materials modeling at the atomic, meso, and continuum scales},
  author = {Thompson, A. and Aktulga, H. and Berger, R. and Bolintineanu, D. and Brown, W. Michael and Crozier, P. and in 't Veld, Pieter J. and Kohlmeyer, Axel and Moore, S. and Nguyen, T. D. and others},
  journal = {Computer physics communications},
  volume = {271},
  pages = {108171},
  year = {2021},
  publisher = {Elsevier},
  doi = {10.1016/j.cpc.2021.108171},
}

@article{Yeh2004,
  title = {System-size dependence of diffusion coefficients and viscosities from molecular dynamics simulations with periodic boundary conditions},
  author = {and, In-Chul Yeh and Hummer, G.},
  journal = {The Journal of Physical Chemistry B},
  volume = {108},
  number = {40},
  pages = {15873--15879},
  year = {2004},
  publisher = {American Chemical Society (ACS)},
  doi = {10.1021/JP0477147},
}

@book{Brunner2017,
  author = {Brunner, Hermann},
  title = {Volterra Integral Equations: An Introduction to Theory and Applications},
  publisher = {Cambridge University Press},
  year = {2017},
  series = {Cambridge Monographs on Applied and Computational Mathematics},
  doi = {10.1017/9781316162491},
}

@article{Persson2017,
  title = {Signatures of solvation thermodynamics in spectra of intermolecular vibrations},
  author = {Persson, Rasmus AX and Pattni, Viren and Singh, Anurag and Kast, Stefan M and Heyden, Matthias},
  journal = {Journal of chemical theory and computation},
  volume = {13},
  number = {9},
  pages = {4467--4481},
  year = {2017},
  publisher = {American Chemical Society (ACS)},
  doi = {10.1021/acs.jctc.7b00184},
}

@book{Zwanzig2001,
  title = {Nonequilibrium statistical mechanics},
  author = {Bellac, M.},
  year = {2007},
  publisher = {Oxford university press},
  journal = {Physics Subject Headings (PhySH)},
  doi = {10.5860/choice.44-5705},
}

@article{Zwanzig1954,
  author = {Zwanzig, Robert W.},
  title = {High-temperature equation of state by a perturbation method. {I}. Nonpolar gases},
  journal = {The Journal of Chemical Physics},
  volume = {22},
  pages = {1420--1426},
  year = {1954},
  doi = {10.1063/1.1740409},
  number = {8},
  publisher = {AIP Publishing},
}

@article{Bennett1976,
  author = {Bennett, Charles H.},
  title = {Efficient estimation of free energy differences from {Monte Carlo} data},
  journal = {Journal of Computational Physics},
  volume = {22},
  pages = {245--268},
  year = {1976},
  doi = {10.1016/0021-9991(76)90078-4},
  number = {2},
  publisher = {Elsevier BV},
}

@article{Beutler1994,
  author = {Beutler, T. and Mark, A. and Schaik, R. V. and Gerber, P. and Gunsteren, W. V.},
  title = {Avoiding singularities and numerical instabilities in free energy calculations based on molecular simulations},
  journal = {Chemical Physics Letters},
  volume = {222},
  pages = {529--539},
  year = {1994},
  doi = {10.1016/0009-2614(94)00397-1},
  number = {6},
  publisher = {Elsevier BV},
}

@article{FrenkelLadd1984,
  author = {Frenkel, Daan and Ladd, Anthony J. C.},
  title = {New {Monte Carlo} method to compute the free energy of arbitrary solids. {Application} to the fcc and hcp phases of hard spheres},
  journal = {The Journal of Chemical Physics},
  volume = {81},
  pages = {3188--3193},
  year = {1984},
  doi = {10.1063/1.448024},
  number = {7},
  publisher = {AIP Publishing},
}

@article{NoyaCondeVega2008,
  author = {Noya, Eva G. and Conde, Maria M. and Vega, Carlos},
  title = {Computing the free energy of molecular solids by the Einstein molecule approach: Ices XIII and XIV, hard-dumbbells and a patchy model of proteins},
  journal = {The Journal of Chemical Physics},
  volume = {129},
  pages = {104704},
  year = {2008},
  doi = {10.1063/1.2971188},
  number = {10},
  publisher = {AIP Publishing},
}

@article{errington2003quantification,
  title={Quantification of order in the Lennard-Jones system},
  author={Errington, Jeffrey R and Debenedetti, Pablo G and Torquato, Salvatore},
  journal={The Journal of chemical physics},
  volume={118},
  number={5},
  pages={2256--2263},
  year={2003},
  publisher={American Institute of Physics}
}

@book{hockney2021,
  title={Computer simulation using particles},
  author={Hockney, Roger W and Eastwood, James W},
  year={2021},
  publisher={crc Press}
}

@article{WCA1971,
  author = {Weeks, John D. and Chandler, David and Andersen, Hans C.},
  title = {Role of repulsive forces in determining the equilibrium structure of simple liquids},
  journal = {The Journal of Chemical Physics},
  volume = {54},
  pages = {5237--5247},
  year = {1971},
  doi = {10.1063/1.1674820},
  number = {12},
  publisher = {AIP Publishing},
}

@article{BarkerHenderson1967,
  author = {Barker, John A. and Henderson, Douglas},
  title = {Perturbation theory and equation of state for fluids. {II}. {A} successful theory of liquids},
  journal = {The Journal of Chemical Physics},
  volume = {47},
  pages = {4714--4721},
  year = {1967},
  doi = {10.1063/1.1701689},
  number = {11},
  publisher = {AIP Publishing},
}

@article{Rosenfeld1977,
  author = {Rosenfeld, Yaakov},
  title = {Relation between the transport coefficients and the internal entropy of simple systems},
  journal = {Physical Review A},
  volume = {15},
  pages = {2545--2549},
  year = {1977},
  doi = {10.1103/PhysRevA.15.2545},
  number = {6},
  publisher = {American Physical Society (APS)},
}

@article{Dzugutov1996,
  author = {Dzugutov, Mikhail},
  title = {A universal scaling law for atomic diffusion in condensed matter},
  journal = {Nature},
  volume = {381},
  pages = {137--139},
  year = {1996},
  doi = {10.1038/381137a0},
  number = {6578},
  publisher = {Springer Science and Business Media LLC},
}

@book{HansenMcDonald,
  author = {Hansen, Jean-Pierre and McDonald, Ian R.},
  title = {Theory of Simple Liquids: With Applications to Soft Matter},
  edition = {4th},
  publisher = {Academic Press},
  year = {2013},
}

@article{Hloucha1999,
  title = {Phase diagram of the four-dimensional Lennard-Jones fluid},
  author = {Hloucha, M. and Sandler, S. I.},
  journal = {J. Chem. Phys.},
  volume = {111},
  number = {17},
  pages = {8043--8047},
  year = {1999},
  doi = {10.1063/1.480138},
}

@article{SantosYusteLopezdeHaro2020,
  title = {Equation of State of Four- and Five-Dimensional Hard-Hypersphere Mixtures},
  author = {L{\'o}pez de Haro, Mariano and Santos, Andr{\'e}s and Yuste, Santos B.},
  journal = {Entropy},
  volume = {22},
  number = {4},
  pages = {469},
  year = {2020},
  doi = {10.3390/e22040469},
}

@article{Wallace1997,
  author  = {Wallace, Duane C.},
  title   = {Statistical mechanics of monatomic liquids},
  journal = {Physical Review E},
  volume  = {56},
  number  = {4},
  pages   = {4179--4186},
  year    = {1997},
  doi     = {10.1103/PhysRevE.56.4179},
}

@book{Wallace2002,
  author    = {Wallace, Duane C.},
  title     = {Statistical Physics of Crystals and Liquids: A Guide to Highly Accurate Equations of State},
  publisher = {World Scientific},
  address   = {Singapore},
  year      = {2002},
}

\section*{Alt Text}
\textbf{Figure~\ref{fig:bias}.} Bar graph of mean signed entropy deviation from the MBWR reference for the Lennard-Jones liquid states. Rigorous hard-sphere 2PT is biased low, Lin-2003 and Desjarlais are biased high, tuned R2PT lies closest to zero, and 3PT partially corrects the deficit, landing between rigorous hard-sphere and the high-biased methods.

\textbf{Figure~\ref{fig:signed}.} Scatter plot of signed entropy deviation from MBWR versus reduced density for Lennard-Jones liquids. Rigorous hard-sphere 2PT lies well below zero, Lin-2003 and Desjarlais above zero, tuned R2PT closest to zero, and 3PT lies below zero by about 0.1 to 0.2, the under-correction largest at the low-density edge.

\textbf{Figure~\ref{fig:metals}.} Two-panel figure comparing liquid-metal entropy errors relative to thermodynamic integration. The left panel shows per-metal deviations and the right panel shows RMS by method; 3PT and tuned R2PT are closest to zero overall.

\textbf{Figure~\ref{fig:cagedos}.} Four-panel stacked decomposition of the density of states into gas, cage, and harmonic-solid components for water and two liquid metals. In water, the rotational cage dominates the librational band, while the translational cage appears at higher frequency than the translational harmonic-solid band.

\textbf{Figure~\ref{fig:lawnm}.} Plot of cage efficiency, the fraction of the hard-sphere entropy deficit that the cage recovers, versus the kernel non-Markovianity for Lennard-Jones liquids, the seven liquid metals, and water. The points follow a universal scaling curve rising from about one-third at low non-Markovianity toward a plateau near one for the strongly caged metals and the water rotational channel; the same Lennard-Jones states sampled coarsely at 64 fs sit higher along the same trend, and water translational channels fall on it without any free-energy reference.

\textbf{Figure~\ref{fig:collapse}.} Plot of the fitted cage prefactor versus inverse dimensionality. The two-dimensional point lies near one-half and the liquid-metal point near one-third, following the predicted trend \(p=1/d\); the monoatomic Lennard-Jones point lies higher (near 0.56) because its weak kernel memory inflates the fitted prefactor, so it is not used to pin the law.

\textbf{Figure~\ref{fig:statefn}.} Two-panel comparison of 3PT entropy differences with MBWR thermodynamic-integration differences for Lennard-Jones liquids. The pairwise differences follow the one-to-one line closely, and the residuals remain narrowly distributed with a small positive offset, indicating a smooth surface with a coherent density-dependent bias rather than random scatter.

\textbf{Table~\ref{tab:lj}.} Table comparing representative Lennard-Jones liquid entropies from MBWR, rigorous hard-sphere 2PT, Lin-2003, Desjarlais, tuned R2PT, and 3PT. The smallest deviations from MBWR are obtained with tuned R2PT; 3PT partially corrects the rigorous hard-sphere deficit, landing short of MBWR on these weakly caged liquids.

\textbf{Table~\ref{tab:metals}.} Table comparing liquid-metal ionic entropies from thermodynamic integration and several 2PT-family methods. 3PT agrees with the reference as closely as the system-tuned R2PT (RMS 0.06 and 0.07 $k_B$ per atom, indistinguishable at seven metals), with no fitted parameter.

\textbf{Table~\ref{tab:water}.} Table of the standard molar entropy of TIP4P/2005 water at 298 K from several methods and reference values. Adding translational and rotational cage terms raises 3PT from the rigorous hard-sphere baseline to near the FEP reference.

\textbf{Table~\ref{tab:waterB}.} Table of the standard molar entropy of SPC/E water at 298 K from several methods and free-energy reference values. The translational and rotational cage terms bring 3PT close to the independently established FEP range.

\end{document}

\title{Supporting Information for: ``An anharmonic liquid-entropy functional from the Mori--Zwanzig memory kernel''}

\author{Ricardo Buarque}
\author{Mauro Gascon}
\affiliation{Program in Materials Science and Engineering, University of California San Diego,
La Jolla, California 92093, USA}
\author{Tod A. Pascal}
 \email{tpascal@ucsd.edu}
\affiliation{Program in Materials Science and Engineering, University of California San Diego,
La Jolla, California 92093, USA}
\affiliation{ATLAS Materials Physics Laboratory, Aiiso Yufeng Li Family Department of Nano and
Chemical Engineering, University of California San Diego, La Jolla, California 92093, USA}

\date{\today}

\maketitle

\tableofcontents

%==============================================================
\section{The \texorpdfstring{$p=1/d$}{p=1/d} Liquid cage prefactor}\label{app:derivation} 
%==============================================================
Here we provide the full derivation of the dimensional-consistency argument of Sec.~\ref{M-sec:theorem}, comprising the per-DoF
normalization, the support-of-the-excess argument that sets $\dpar$, and the gas mirror, together with
the dynamical/projection-operator reading and its direct numerical test. We also
state explicitly the one step that resists a closed form.

\paragraph{Per-degree-of-freedom argument.} The kernel is
inverted from the normalized trace VACF $c(t)=c_\perp(t)$, which is channel-independent by
isotropy, so $\FK,\FM,\gamma$ and the weights $W_s,\;W_g=s_{\rm HS,d}/d$ are per-DoF; the
per-particle cage entropy is the single-channel re-weight, $=\tfrac1d\times$ trace.

\paragraph{Gas--cage entropy asymmetry from 2nd order perturbation theory} The gas accesses the full $d$-dim hard-sphere entropy
through $d$ free-translation channels ($\int S^{\rm gas}W_g=f\,s_{\rm HS,d}$). This has no $1/d$ dependence. The cage is thus 
one bounded state per particle, and its excluded-volume excess is counted once, corresponding to the $d_\parallel = 1$ radial coordinate from the main text. That single count is the $\ell=0$ monopole of the cage, and since a molecular cage is anisotropic, one needs to
account for the angular content explicitly. To do this correctly, two distinct cases must be considered.

(i) The ensemble velocity-correlation tensor
$\langle v_\alpha(0)\,v_\beta(t)\rangle$, a dynamical object from which the cage spectrum is built. In an isotropic liquid this exactly equates to 
$\tfrac13\delta_{\alpha\beta}\,C(t)$ by rotational invariance of the equilibrium ensemble (i.e., exactly equal to $\ell=0$). It is captured completely by the trace VACF, since the rank-2 tensor has no traceless
($\ell=2$) component by liquid isotropy. The instantaneous tensor does carry $\ell=2$, but it averages to zero in the lab frame that defines
the VACF. 

(ii) The configurational angular anisotropy of the instantaneous cage, as a correction to the above. In a perfect tetrahedron, $\ell=1,2$ vanish identically and the first survivor is the $\ell=3$ $T_d$ invariant. One needs not make this assumption however. Instead, we
measure this quantity directly from our simulation, spanning $69{,}120$ molecular environments over $40$ frames with four nearest neighbors. We find that the first-shell oxygen geometry of TIP4P/2005 at $298$~K is, unsurprisingly, a
distorted tetrahedron, with an Errington--Debenedetti order parameter~\cite{errington2003quantification} $q_{\rm tet}=0.67$, not $1$. This means that while $\ell=2$ is indeed suppressed, it is not exactly zero. Its bond-order angular power, $P(\ell{=}2)/P(0)=0.10$, is an order of magnitude below the monopole and one-third of the $T_d$-allowed $\ell=3$ channel ($P(3)/P(0)=0.32$). 

We argue that $p=1/d$ reproduces the water entropy
without modification, not because $\ell=2$ is symmetry-forbidden (in the real, distorted cage we now know that it is
merely suppressed), but because the dynamical cage tensor is exactly $\ell=0$ by isotropy and
the measured configurational $\ell=2$ correction is second-order and small.

For a pure anisotropic perturbation of the cage, \begin{equation} U_\epsilon(\bm{x}) = U_0(r) + \epsilon\,u(r)Y_{\ell}(\Omega), \qquad \ell \geq 1, \end{equation} the first-order response of the excess entropy vanishes by rotational invariance of the unperturbed isotropic ensemble. Every first-order term contains the angular average \begin{equation} \left\langle Y_{\ell}\right\rangle_0 = \frac{1}{4\pi} \int_{\mathbb{S}^2} Y_{\ell}(\Omega)\,\mathrm{d}\Omega = 0, \qquad \ell \geq 1, \end{equation} because every nonmonopolar spherical harmonic is orthogonal to the constant mode. Thus, the entropy of the anisotropic cage and the Einstein frequency that fixes the harmonic reference are separately stationary at $\epsilon=0$. The residual correction enters at second order in $\epsilon$, with a positive, computable coefficient. We compute these coefficients across our material systems, and find that the metals' $p^*=0.37$ corresponds to sub-percent anisotropic power in
the icosahedral $\ell=6$ channel. The agreement of $p=1/3$ with the
water FEP benchmarks bounds $\Pi_2\lesssim0.02$, an order of
magnitude below the geometric bond-order ratio $P(2)/P(0)=0.10$ measured
above. This separation is consistent with configurational averaging in the potential of mean force: the latter integrates over the unresolved thermal and many-particle degrees of freedom, whereas $P(2)/P(0)$ characterizes a single four-neighbor geometry. The two measures therefore need not have comparable magnitudes.

\paragraph{The prefactor is fixed by cage symmetry.} Consider the configurational entropy of a
tagged particle in a confining potential, relative to the isotropic-harmonic reference. For
separable ("box") confinement, $U(\bm x)=\sum_\alpha u(x_\alpha)$, the $d$ Cartesian
coordinates are independent: the excess entropy is $d$ copies of the 1D excess, the cage spectrum is
$\mathrm{cage}=d\,\mathrm{cage}_\perp$, and consistency requires $p=1$. For isotropic
confinement, $U=U(r)$ with $r=|\bm x|$, the configurational integral factorizes as
$Z=\Omega_{d-1}\!\int_0^\infty e^{-\beta U(r)}r^{d-1}dr$; the angular surface $\Omega_{d-1}$ is
identical for the anharmonic potential and its harmonic reference and cancels in the
excess. The entropy that distinguishes a real cage from a harmonic well is carried by the single
radial coordinate while the cage spectrum still superposes all $d$ channels, giving $p=1/d$.
Formally, $p$ tracks the symmetry of the confinement: $p=1$ for separable (box-like)
confinement and $p=1/d$ for isotropic confinement, the two coinciding at $d=1$. 

\paragraph{The radial Jacobian contributions to $1/d$.} We note that only the
angular constant $\Omega_{d-1}$ cancels in the excess; the radial Jacobian $r^{d-1}$ remains
inside both integrals, and the radial excess
$\ln\!\int e^{-\beta U}r^{d-1}dr-\ln\!\int e^{-\beta U_{\rm harm}}r^{d-1}dr$ is an explicit function
of $d$ (its moments are $\langle r^n\rangle=2^{n/2}\Gamma((d{+}n)/2)/\Gamma(d/2)$). It might appear that this surviving $d$-dependence corrupts the projection factor $1/d$. However, this is not the case: the construction never explicitly evaluates the configurational radial integral, and the Jacobian's $d$-dependence is not
dropped but relocated into the reference weight. Practically, this means that the cage spectrum is weighted
by $(W_g-W_s)$, the per-DoF difference between a hard-sphere gas reference and a harmonic solid
reference, and $W_g$ is the full $d$-dimensional hard-sphere entropy per degree of freedom:
$W_g=\tfrac1d\bigl[\tfrac{d}{2}+1+\ln(\Lambda^{-d}\,v/f)\bigr]+\tfrac1d\,s^{\rm ex}_{\rm HS}(y)$, carrying
the $d$-dimensional kinetic phase space ($\Lambda^{-d}$) and the $d$-dimensional Carnahan--Starling
excess $s^{\rm ex}_{\rm HS}(y)$. These are exactly the radial-measure ($r^{d-1}$) content of the
configurational integral; the gas reference is the $d$-dimensional hard sphere.

The prefactor is then only the angular/per-DoF projection count: $\Omega_{d-1}$ cancels and the velocity
projection $\langle\bm v\!\cdot\!\bm v\rangle=d\,k_BT/m$ normalizes per DoF, so the cage, one
longitudinal channel of the $d$ equivalent velocity channels in flat, Jacobian-free velocity space
(the dynamical view below), contributes $1/d$. The two $d$-dependencies are thus separated by
construction: the angular projection gives $p=1/d$, while the radial measure lives in the
$d$-dimensional gas weight $W_g$, not in $p$. This separation is evidenced by the data. Where the cage is on its saturated plateau, so that the
data-optimal $p^*$ relates to the prefactor rather than the memory shortfall of Sec.~\ref{M-sec:lawnm}. It
is $p^*=0.37\pm0.06$ for the $d=3$ metals and $0.51$ for the $d=2$ LJ liquid: $1/d$ to within
uncertainty, bounding any residual leakage of the radial measure into $p$ to $\lesssim0.04$ at $d=3$. A radial
correction at or below this $\lesssim0.04$ bound cannot be excluded, but would not be dominant to the
projection. In any case, it would be absorbed by the hard-sphere reference rather than left as a free
constant, so the "parameter-free" status of $p=1/d$ stands.

\paragraph{The configurational and dynamical pictures are one mechanism, not two.} We argue that the
cage-symmetry argument above is
the configurational statement of a counting, the same counting as the Mori projection in velocity
space. Projecting onto the velocity, $\mathcal P A=(\langle A\!\cdot\!\bm v\rangle/\langle\bm
v\!\cdot\!\bm v\rangle)\bm v$, carries $\langle\bm v\!\cdot\!\bm v\rangle=d\,\kB T/m$, so the kernel
($K(t)=\langle\bm R(0)\!\cdot\!\bm R(t)\rangle/\langle\bm v^2\rangle$, $\bm R$ the projected force of
Sec.~\ref{M-sec:cage}) and the responses $\FK,\FM$ are intrinsically per-degree-of-freedom. The
cage, i.e., longitudinal backscattering--the momentum reversal that makes the negative VACF dip, is the
single velocity channel along the instantaneous direction of confinement, which by isotropy is the
same degree of freedom as the radial coordinate of the cage-symmetry argument. The two are thus not independent
derivations that happen to coincide but the position-space and velocity-space descriptions of
identifying the one excess-bearing channel of $d$. 

\paragraph{The memory fraction is the integrand, not the prefactor.} As a point of emphasis, we note that the cage spectrum is built from the memory excess $\FK-\FM$. As a result, one might
expect the prefactor to track the kernel's memory content, i.e., the band-averaged response memory
fraction $\langle r\rangle_g=\langle(\FK-\FM)/\FK\rangle_g$, which is $O(1)$ ($\approx0.63$, very near to $2/3$ in three dimensions). We argue that this is not the case, precisely because the
spectrum already carries the memory. The cage entropy is $p\!\int g\,d\nu$ with
$g=\mathrm{cage}\,(1-w)(W_g-W_s)$ and $\mathrm{cage}\propto\FK-\FM$, i.e., the integrand is the
memory excess. Re-weighting it by a frequency-dependent memory fraction $r(\nu)$ gives $\Delta
S=\langle r\rangle_g\!\int g$, which counts the memory twice: once in the spectrum and once in
the weight, and over-weights the cage by $\langle r\rangle_g/(1/d)\approx1.9$. This overcounting would predict $0.63$ vs
$1/3$, roughly doubling it. The correct prefactor must therefore be
memory-independent: a frequency-flat count of how many degrees of freedom the gas-versus-solid
distinction is assigned to, fixed by the cage-symmetry argument at $1/d$.

\paragraph{The entropy excess tracks the projection count and not the memory.} The efficiency
equation of Sec.~\ref{M-sec:lawnm} is the practical realization of the above split. The system-dependent memory lives
in the cage spectrum, so $\Delta S_{\rm cage}$ and the efficiency $\eta_{\rm cage}$ vary with the
degree of non-Markovianity $\gamma/\Omega_0$. The prefactor stays the fixed projection count $1/d$. If $p$ was
instead the memory fraction, the memory would enter twice and $\eta_{\rm cage}$ would scale with it
super-linearly. This would mean that the LJ, metal and water states could not collapse onto one curve. They do
collapse, with a flat $1/d$ multiplying a memory-dependent cage, which in itself is direct evidence that $1/d$ is
the projection normalization while the memory is the integrand. In summary, the entropy excess tracks the count
of excess-bearing channels (one of $d$), not the magnitude of the kernel's memory, which is instead supplied by the spectrum.

\paragraph{Dimensionality argument in 4D.} The four-dimensional liquid provides a stringent test of the proposed
prefactor, but it is not well suited to a direct fitted-$p$ analysis. In $d=4$,
the soft-mode anharmonic deficit is substantially larger than in $d=3$ and
therefore contaminates the data-optimal prefactor $p^*$ obtained by simply
centering $S_{\rm 3PT}$ on the thermodynamic-integration reference. This effect
inflates $p^*$ to values of order unity, even for strongly caged states, and
therefore obscures the dimensional projection factor itself.

We instead test whether a candidate prefactor leaves a physically reasonable
non-memory residual after the memory-derived cage has been accounted for. Since
$\Delta S_{\rm cage}(p)$ is linear in $p$, each choice of $p$ implies a residual
soft-mode anharmonic entropy per mode,
\[
\Delta s_{\rm anh}(p)=
\frac{1}{d}
\left[
(S_{\rm TI}-S_{\rm rigHS})
-
\frac{\Delta S_{\rm cage}(p)}{\eta(\mathcal{M})}
\right],
\]
where
\[
\eta(\mathcal{M})=1.05\left(1-e^{-2.386\,\mathcal{M}}\right)
\]
is the lower-dimensional efficiency curve expressed in terms of the
weight-free memory area
\[
\mathcal{M}=\frac{\int |F_K-F_M|\,d\nu}{\int |F_K|\,d\nu}.
\]
The asymptote slightly above unity is an artifact of the unconstrained fit
(the lower-dimensional data do not reach the $\mathcal{M}$ plateau, so the
asymptote is weakly determined); the physical bound is $\eta\to1$, and the
constrained refit reported below leaves the conclusion of this test
unchanged. A consistent prefactor should leave a residual that is positive and nearly
state-independent, because the remaining soft-mode correction is not expected
to scale with the cage memory.

\begin{figure}[tbp]
\includegraphics[width=\columnwidth]{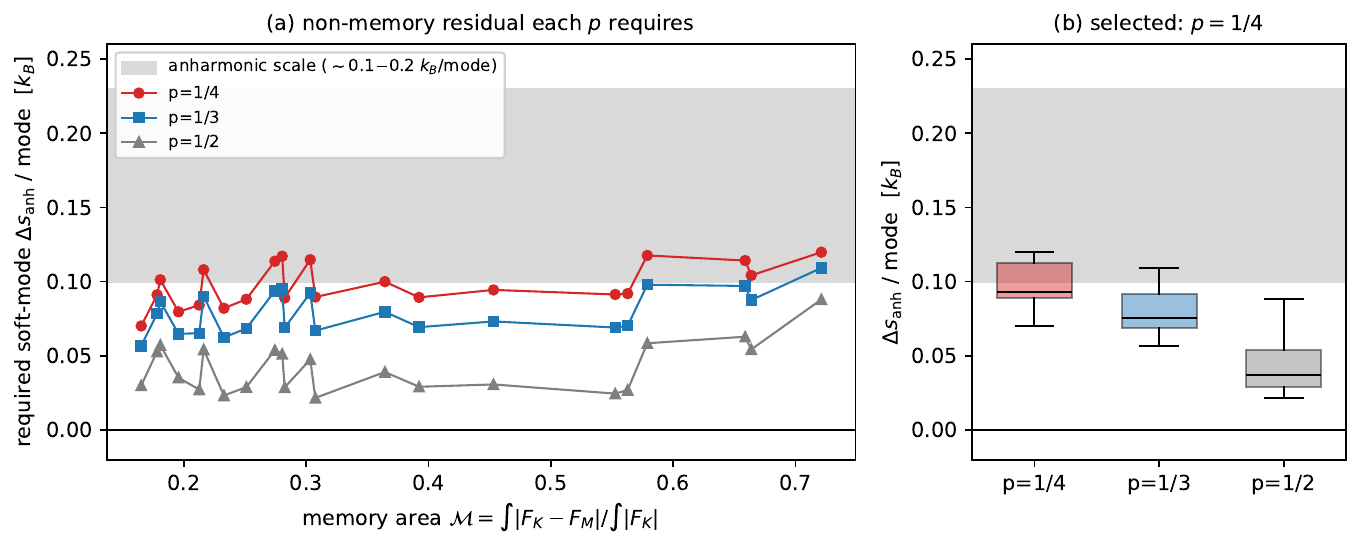}
\caption{\label{fig:d4prefactor}\textbf{The $d=4$ cage-prefactor test.}
(a)~Residual non-memory anharmonic entropy per mode implied by candidate
prefactors $p\in\{1/4,1/3,1/2\}$ after the memory-derived cage contribution is
placed on the lower-dimensional efficiency curve $\eta(\mathcal{M})$. Results
are shown for $22$ four-dimensional LJ fluid states. The choice $p=1/4$ leaves a
positive, nearly state-independent residual of
$\simeq0.10\,\kB$/mode; $p=1/3$ is slightly lower and mildly sloped; $p=1/2$
over-subtracts the cage contribution and gives a small, memory-dependent
residual. The shaded region indicates the expected scale of per-mode soft
anharmonic entropy. (b)~Residual distributions for each prefactor
(box~$=$~IQR, whiskers~$=$~range).}
\end{figure}

Figure~\ref{fig:d4prefactor}(a) applies this test to $22$ four-dimensional LJ
fluid states, including the $T^*=3.0$ and $4.0$ isotherms and the $P^*=3.2$
isobar. The choice $p=1/4$ leaves a flat residual of
$\Delta s_{\rm anh}\simeq0.10\,\kB$ per mode, with a coefficient of variation of
$14\%$. This scale is consistent with the expected per-mode soft anharmonic
entropy. The choice $p=1/3$ gives a slightly smaller residual and a mild
systematic slope. By contrast, $p=1/2$ over-subtracts the memory contribution:
the residual falls to approximately $0.04\,\kB$ per mode and develops a much
larger, memory-dependent scatter, especially for the strongly caged states.
Figure~\ref{fig:d4prefactor}(b) summarizes these distributions.

This conclusion is robust to the calibration of $\eta(\mathcal{M})$:
constraining the asymptote to its physical bound $\eta_\infty\le1$ (refit,
$\eta(\mathcal{M})=1-e^{-1.44\,\mathcal{M}}$) sharpens rather than weakens the
discrimination ($p=1/4$: mean residual $0.069\,\kB$/mode, CV $24\%$; $p=1/3$:
CV $46\%$; $p=1/2$: over-subtracts to a negative mean residual). Thus the four-dimensional data support $p=1/4$ over $p=1/2$ and favor it over
$p=1/3$, completing a consistency-level dimensional sequence
$p=1/d$ for $d=2,3,4$. This argument is not a standalone analytic proof: it uses
the lower-dimensional efficiency curve to separate memory and non-memory
anharmonicity. A direct four-dimensional model of the soft-mode residual would
make this test sharper, but it outside the scope of the current work.

\paragraph{Limitations and future work.} A fully analytic proof would require computing the cage spectrum (from
the VACF) and the cage entropy (the radial excess above) for one model, and show they agree
with $p=1/d$. This requires a system with simultaneously known equation-of-state entropy and
analytic VACF that possesses both a gas and a cage phase, which dense liquids do not afford in closed
form: single-particle confined models have known entropy but no gas phase, $f\to0$, where $W_g$
diverges. We therefore close the derivation with the cage-symmetry argument, exact in the box
limit and reducing the isotropic case to the single radial coordinate, together with the numerical
confirmation at $d=2$ ($\tfrac12$) and $d=3$ ($\tfrac13$), and the $d=4$ test below ($\tfrac14$). Formularizing the derivation along the prescribed lines will be the subject of a follow-up work.

%==============================================================
\section{Practical considerations}
\label{app:scope}
%==============================================================
This section clarifies two assumptions behind the cage construction of
Secs.~\ref{M-sec:cage}--\ref{M-sec:cageentropy}: why the cage is reweighted with
an excluded-volume entropy reference, and when the sequential gas--cage
decomposition is expected to be valid.

\paragraph{Hard-sphere weight as an entropy reference.}
The weight $W_g$ is an entropy reference, not a microscopic model of the cage
potential. A cage collision in a soft liquid is neither a hard-sphere collision
nor a harmonic vibration. The use of $W_g$ rests instead on the standard
liquid-state result that dense-liquid entropy is largely controlled by excluded
volume: perturbative treatments such as Weeks--Chandler--Andersen and
Barker--Henderson theory place the structural and entropic role mainly in the
repulsive core, while attractions primarily shift the energy
\cite{WCA1971,BarkerHenderson1967,HansenMcDonald}. Excess-entropy scaling gives
the same message, collapsing many continuous potentials onto hard-sphere-like
entropy trends \cite{Rosenfeld1977,Dzugutov1996}.

Thus the cage is not assumed to be a hard-sphere well. Rather, the
non-Markovian cage band is assigned the same excluded-volume reference that 2PT
already uses for its diffusive gas. The harmonic weight $W_s$ and the
hard-sphere weight $W_g$ are the two limiting entropy references, and
$(W_g-W_s)$ measures how much entropy is gained by treating the bounded,
diffusively coupled cage band as a liquid-like excluded-volume motion rather
than as a harmonic oscillator. The fact that this same reweighting works without
retuning across Lennard-Jones liquids, Sutton--Chen metals, and water is the
empirical evidence that the correction captures a transferable excluded-volume
component of the cage entropy.

\paragraph{Scope of the sequential gas--cage extraction.}
The present construction is sequential. The Markovian gas, fixed by the
zero-frequency friction $\gamma=\tilde K(0)$, is removed first; the cage is then
defined as the positive non-Markovian memory excess over this gas reference.
Consequently $\mathrm{cage}(0)=0$, and all zero-frequency diffusive weight is
assigned to the gas. This decomposition is appropriate when diffusion and cage
rattling are spectrally separated, as in the equilibrium liquids studied here.

The same separation need not hold in deeply supercooled or otherwise slowly
relaxing liquids. When structural ($\alpha$) relaxation and cage ($\beta$)
relaxation overlap, a single-$\gamma$ Markovian gas can misassign low-frequency
non-Markovian relaxation. In that regime, a simultaneous gas--cage
decomposition, for example using a multi-mode memory kernel or a
mode-coupling-type closure, may be more appropriate.

\paragraph{What the cage responds to.}
The construction is selective by design: the cage is fed only by
non-Markovian spectral structure below the Einstein cutoff, which in a
liquid is generated by escape-coupled (structural-relaxation) anharmonicity;
harmonic memory and bound anharmonic motion contribute essentially nothing,
so the cage correctly ignores spectral weight that is bound (crystals,
molecular vibrations) or freely diffusive (gases).

%==============================================================
\section{Rotational 3PT for rigid molecules: kernel, weights, and prefactor} \label{app:rot3pt}
%==============================================================

For rigid molecular liquids, the rotational contribution can be treated in close
analogy with the translational 3PT construction. The required changes are the
use of an inertia-weighted angular-velocity autocorrelation, a free-rigid-rotor
entropy reference for the gas-like rotational channel, and a dimensional
projection factor based on the number of active rotational degrees of freedom.
The construction below is the one used for the rigid water models in
Sec.~\ref{M-sec:water}. 

\paragraph{Rotational density of states.}
For a rigid molecule with angular velocity \(\bm\omega(t)\) in the body frame and principal moments of inertia \(I_\alpha\), we form the mass-weighted rotational autocorrelation
\begin{equation}
C_{\rm rot}(t)
=
\sum_{j}\sum_{\alpha=1}^{d_{\rm rot}}
I_{j,\alpha}\,
\big\langle \omega_{j,\alpha}(0)\,\omega_{j,\alpha}(t)\big\rangle ,
\end{equation}
with \(d_{\rm rot}=3\) for a nonlinear rigid molecule and \(d_{\rm rot}=2\) for a linear rigid molecule. Its cosine transform defines the rotational density of states,
\begin{equation}
S_{\rm rot}(\nu)
=
\frac{4}{k_{B}T}\int_{0}^{\infty} C_{\rm rot}(t)\cos(2\pi \nu t)\,dt,
\qquad
\int_{0}^{\infty} S_{\rm rot}(\nu)\,d\nu
=
d_{\rm rot}N .
\label{eq:Srot}
\end{equation}
Thus the rotational DoS is normalized exactly as in translation, but over the \(d_{\rm rot}\) rotational channels rather than the \(d\) Cartesian ones. In the water calculations of Sec.~\ref{M-sec:water}, \(d_{\rm rot}=3\).

\paragraph{Rotational Mori kernel and the cage spectrum.}
Let
\begin{equation}
c_{\rm rot}(t)=\frac{C_{\rm rot}(t)}{C_{\rm rot}(0)}
\end{equation}
be the normalized rotational autocorrelation. The rotational Mori--Zwanzig equation has the same Volterra form as the translational one,
\begin{equation}
\dot c_{\rm rot}(t)
=
-\int_{0}^{t} K_{\rm rot}(t-t')\,c_{\rm rot}(t')\,dt' ,
\label{eq:rot_volterra}
\end{equation}
which we invert in the time domain to obtain the rotational memory kernel \(K_{\rm rot}(t)\). Its half-sided Fourier transform \(\tilde K_{\rm rot}(\omega)\) defines the zero-frequency rotational friction
\begin{equation}
\gamma_{\rm rot}\equiv \tilde K_{\rm rot}(0).
\end{equation}
As in Sec.~\ref{M-sec:cage}, two responses bracket the dynamics:
\begin{align}
F_{K}^{\rm rot}(\nu)
&=
{\rm Re}\!\left[\frac{1}{i\omega+\tilde K_{\rm rot}(\omega)}\right],\\
F_{M}^{\rm rot}(\nu)
&=
\frac{\gamma_{\rm rot}}{\gamma_{\rm rot}^{2}+\omega^{2}},
\end{align}
with \(\omega=2\pi c\nu\). The rotational cage spectrum is the non-Markovian excess over the memoryless response,
\begin{equation}
\mathrm{cage}_{\rm rot}(\nu)
=
\mathrm{clip}\!\Big[
\kappa_{\rm rot}\big(F_{K}^{\rm rot}(\nu)-F_{M}^{\rm rot}(\nu)\big),
\,0,\,
S_{{\rm rot},2}^{\rm sol}(\nu)
\Big],
\label{eq:cagerot}
\end{equation}
where \(\kappa_{\rm rot}\) fixes the amplitude in DoS units and
\(S_{{\rm rot},2}^{\rm sol}=\max(S_{\rm rot}-S_{\rm rot}^{\rm gas},0)\)
bounds the available solid weight. By construction,
\(\mathrm{cage}_{\rm rot}(0)=0\),
so the rotational cage carries no spurious free-rotor weight at zero frequency. This is the rotational counterpart of the translational cage of Sec.~\ref{M-sec:cage}.

\paragraph{Rotational gas and solid weights.}
The rotational gas reference is the classical free rigid rotor, not a
hard-sphere fluid. Rotational diffusion does not carry a translational
excluded-volume entropy, so the gas-like rotational weight is the per-degree-of-
freedom rigid-rotor entropy \(W_g^{\rm rot}=S^R/(k_B d_{\rm rot})\). For a
nonlinear molecule,
\[
\frac{S^R}{k_B}
=
\frac{3}{2}
+
\ln\!\left[
\frac{\pi^{1/2}}{\sigma}
\left(
\frac{T^3}{\theta_A\theta_B\theta_C}
\right)^{1/2}
\right],
\]
with the usual principal rotational temperatures and symmetry number \(\sigma\).
The harmonic weight \(W_s(\nu)\) is retained for the solid-like librational
remainder.

The use of \(W_g^{\rm rot}\) for the cage correction should be interpreted as a
reference-state assignment, not as a claim that the librational cage is
dynamically free-rotor-like. It replaces the harmonic entropy of a hindered,
finite-memory rotational band by the entropy reference appropriate to the
rotational diffusive limit.

\paragraph{Rotational projection prefactor.}
The rotational prefactor follows the same trace normalization logic used for
translation. The kernel \(K_{\rm rot}(t)\), the friction \(\gamma_{\rm rot}\),
and the responses \(F_K^{\rm rot}\) and \(F_M^{\rm rot}\) are obtained from the
normalized rotational autocorrelation and are therefore per rotational degree of
freedom. The rotational DoS entering the entropy integral, however, is traced
over \(d_{\rm rot}\) angular channels. For an isotropically averaged liquid
rotor, we therefore assign the cage reweighting to one projected rotational
channel out of the \(d_{\rm rot}\) channels in the trace:
\[
p_{\rm rot}=\frac{1}{d_{\rm rot}}.
\]
Thus \(p_{\rm rot}=1/3\) for nonlinear rigid water and \(p_{\rm rot}=1/2\) for a
linear rigid molecule. As in translation, this factor is a projection count; the
frequency-dependent memory content is already contained in
\(\mathrm{cage}_{\rm rot}(\nu)\).

\paragraph{Three-phase rotational entropy.}
The full rotational entropy is therefore
\begin{equation}
S_{\rm 3PT}^{\rm rot}
=
\underbrace{\int S_{\rm rot}^{\rm gas}W_{g}^{\rm rot}\,d\nu}_{S_{\rm gas}^{\rm rot}}
+
\underbrace{\int \mathrm{cage}_{\rm rot}\Big[W_{s}
+p_{\rm rot}\,g(f_{\rm rot})(1-w_{\rm rot})(W_{g}^{\rm rot}-W_{s})\Big]\,d\nu}_{S_{\rm cage}^{\rm rot}}
+
\underbrace{\int S_{\rm rot}^{\rm sol}W_{s}\,d\nu}_{S_{\rm solid}^{\rm rot}},
\label{eq:S3PTrot}
\end{equation}
with
\(
S_{\rm rot}
=
S_{\rm rot}^{\rm gas}
+
\mathrm{cage}_{\rm rot}
+
S_{\rm rot}^{\rm sol}
\). In liquid water, this construction reclassifies the hindered libration band from harmonic-solid to caged free-rotor motion, yielding the dominant molecular correction reported in Sec.~\ref{M-sec:water}.

\paragraph{Low-frequency structure of the rotational excess (verification).}
For water the rotational channel is strongly overdamped relative to the
librational frequency range. In the TIP4P/2005 trajectory at 298~K, for
example, \(\gamma_{\rm rot}\) is larger than the width of the rotational
spectrum, so the Markovian response is nearly flat over the librational band.
The positive excess \(F_K^{\rm rot}-F_M^{\rm rot}\) then identifies the
librational cage, while the compensating negative branch lies primarily below
the libration band and is removed by the clipping step. The zero-sum identity of
the unclipped response is therefore preserved at the response level, even though
only the positive excess is assigned to the explicit cage phase.

This behavior also marks the main limitation of the current rotational closure:
for a nearly arrested rotational channel, a free-diffusion Markovian rotor is a
weak null model. A natural refinement would be to replace \(F_M^{\rm rot}\) by a
parameter-free damped-librator or single-exponential-memory GLE reference
matched to the exact kernel invariants \(K(0)=\Omega_0^2\) and
\(\tilde K(0)=\gamma\). Such a reference would preserve the zero-frequency response,
\(F_{\rm ref}(0)=F_K(0)=1/\gamma\), while replacing the freely diffusive
Markovian baseline by a librational one. We leave this extension for
future work.

\section{Self-solvation free energy by two-leg deletion FEP}\label{app:fep}

The independent free-energy reference for the liquid entropy is the
self-solvation free energy of one water molecule in its own liquid, i.e., the
excess chemical potential \(\mu_{\rm ex}\) at the simulated liquid density and
temperature. We compute \(\mu_{\rm ex}\) by deleting a tagged molecule from the
bath through a two-leg free-energy perturbation (FEP) path in LAMMPS
\cite{Zwanzig1954,Plimpton1995}. The first leg discharges the tagged molecule:
with Lennard-Jones interactions fully coupled, the partial charges are scaled as
\(q_i(\lambda)=(1-\lambda)q_i^{(0)}\). The second leg decouples the Lennard-Jones
interactions of the now-neutral tagged molecule using a soft-core potential,
which avoids the end-point singularity of linear Lennard-Jones scaling
\cite{Beutler1994}. With this deletion convention,
\[
\mu_{\rm ex}=-(\Delta F_q+\Delta F_{\rm vdw}),
\]
where \(\Delta F_q+\Delta F_{\rm vdw}\) is the free energy for decoupling the
tagged molecule from the liquid.

Free-energy differences between neighboring windows are evaluated with the
Bennett acceptance ratio (BAR) \cite{Bennett1976}; direct exponential averages
are computed as convergence checks. The \(\lambda\) ladders are chosen from
short pilot runs to approximately equalize the perturbation variance across
windows, typically requiring 11--16 windows per leg. Each window is equilibrated
for \(0.1\) ns and sampled for \(0.4\) ns at \(298\) K and \(1\) atm. The bath
contains \(N=1728\) water molecules, matching the trajectories used for the
2PT/3PT analysis.

\subsection{Explicit M-site treatment for TIP4P/2005}

For TIP4P-family models, the negative charge resides on the massless M-site
rather than on the oxygen. In the implicit \texttt{pppm/tip4p} treatment in
LAMMPS, this M-site is reconstructed geometrically from the oxygen and hydrogen
positions at each step. Therefore, scaling the oxygen charge does not scale the
M-site charge and would leave the discharge leg incomplete. To avoid this, the
tagged TIP4P/2005 molecule is represented explicitly as a four-site rigid
molecule: the M-site is assigned its own atom type, charge, and a small mass
(\(10^{-7}\) amu), and the tagged molecule is kept rigid with
\texttt{fix rigid/small}. The M-site charge can then be scaled directly by atom
type during the discharge leg.

This explicit representation reproduces the potential energy of the corresponding
implicit TIP4P/2005 configuration to within \(0.055\%\). The van der Waals energy
is unchanged to within \(0.002\%\); the remaining difference is purely
electrostatic. SPC/E has all charges on real atoms and therefore requires no
analogous treatment.

\subsection{From \texorpdfstring{\(\mu_{\rm ex}\)}{mu_ex} to liquid entropy}

The excess chemical potential is combined with the ideal translational and
rotational entropies of the isolated rigid molecule and the mean intermolecular
potential energy of the liquid to obtain
\begin{equation}
S^\circ =
S^{\rm ideal}_{\rm trans}(T,\rho_{\rm liq})
+
S^{\rm ideal}_{\rm rot}(T)
+
\frac{\langle U_{\rm inter}\rangle-\mu_{\rm ex}}{T}
-
R .
\label{eq:Sfep}
\end{equation}
Here \(S^{\rm ideal}_{\rm trans}\) is evaluated at the liquid number density, and
\(S^{\rm ideal}_{\rm rot}\) is the rigid-rotor entropy with symmetry number
\(\sigma=2\). The final \(-R\) term is the residual-pressure correction. FEP
returns the excess chemical potential at fixed liquid density, whereas the
residual entropy is obtained from the excess Helmholtz free energy. At liquid
density \(P-\rho k_BT\simeq-\rho k_BT\), so \(A_{\rm ex}=\mu_{\rm ex}+RT\);
omitting this term would overestimate \(S^\circ\) by \(R=8.31\)
J\,mol\(^{-1}\)K\(^{-1}\).

\begin{table}[t]
\caption{\label{tab:fep}Assembly of the FEP self-solvation entropy
[Eq.~\eqref{eq:Sfep}] for liquid TIP4P/2005 and SPC/E water at \(298\) K.}
\begin{ruledtabular}
\begin{tabular}{lccl}
Quantity & TIP4P/2005 & SPC/E & source \\
\colrule
\(S^{\rm ideal}_{\rm trans}\) (J\,mol\(^{-1}\)K\(^{-1}\)) & 84.93 & 84.94 & Sackur--Tetrode, \(\rho_{\rm liq}\) \\
\(S^{\rm ideal}_{\rm rot}\) (J\,mol\(^{-1}\)K\(^{-1}\)) & 43.70 & 44.59 & rigid rotor, \(\sigma=2\) \\
\(\langle U_{\rm inter}\rangle\) (kcal\,mol\(^{-1}\)) & \(-11.35\) & \(-11.09\) & 1728-molecule MD \\
\(\mu_{\rm ex}\) (kcal\,mol\(^{-1}\)) & \(-7.19\pm0.04\) & \(-7.09\pm0.03\) & two-leg deletion BAR, five seeds \\
\(S^\circ_{\rm FEP}\) (J\,mol\(^{-1}\)K\(^{-1}\)) & \textbf{61.9\(\pm\)0.5} & \textbf{65.1\(\pm\)0.4} & Eq.~\eqref{eq:Sfep} \\
\end{tabular}
\end{ruledtabular}
\end{table}

For SPC/E, the computed \(\mu_{\rm ex}=-7.09\pm0.03\) kcal\,mol\(^{-1}\)
(five independent seeds; SEM quoted) is close to
independent literature values for the same model:
\(-7.0\) kcal\,mol\(^{-1}\) from regularized potential-distribution theory
\cite{WeberMerchantAsthagiri2011}, \(-6.9\) kcal\,mol\(^{-1}\) from
quasi-chemical theory \cite{MerchantAsthagiri2011}, and
\(-7.05\) kcal\,mol\(^{-1}\) from BAR-FEP \cite{ShirtsPande2005} (the closest to
our value). These values
correspond to \(S^\circ_{\rm FEP}\) in the range \(62\)--\(65\)
J\,mol\(^{-1}\)K\(^{-1}\), while our own FEP value gives
\(65.1\pm0.4\) J\,mol\(^{-1}\)K\(^{-1}\). The 3PT values reported in the main text
fall within the uncertainty implied by these free-energy references, with no
2PT or 3PT partition entering the FEP route.

\section{Absolute entropy by Frenkel--Ladd / Einstein-molecule TI}
\label{app:ti}

As a second independent absolute-entropy check for TIP4P/2005, we compute the
entropy of ice Ih by Frenkel--Ladd thermodynamic integration in the
Einstein-molecule formulation of Noya, Conde and Vega
\cite{FrenkelLadd1984,NoyaCondeVega2008}. The resulting crystal entropy is then
bridged to the liquid at \(298\) K. Because this route involves a small
difference of large free-energy terms and a melting/heat-capacity bridge, we use
it as a consistency check rather than as the primary liquid benchmark.

\paragraph*{Crystal free energy.}
We couple a proton-disordered ice-Ih cell of \(N=432\) rigid TIP4P/2005
molecules to an Einstein-molecule reference at \(252\) K, near the model melting
point. Each molecule is tethered to its lattice position and orientation by
harmonic springs (\texttt{fix spring/self}) with a stiffness
\(K=2500\) kcal mol\(^{-1}\)\AA\(^{-2}\). The coupling work,
\[
W_{\rm spring}
=
\int_0^1
\left\langle
\frac{\partial U_{\rm spring}}{\partial \lambda}
\right\rangle
d\lambda,
\]
is evaluated over a 32-point \(\lambda\) ladder, giving
\(W_{\rm spring}=29.12\) kJ mol\(^{-1}\).

The reference lattice is an energy-minimized configuration with
\(U_{\rm min}=-63.07\) kJ mol\(^{-1}\) per molecule. Using a finite-temperature
snapshot instead would incorrectly include thermal distortion in the lattice
binding energy. The correction from the ideal Einstein field to the interacting
solid is evaluated as
\[
\Delta A_1 =
U_{\rm min}
-
k_B T
\ln
\left\langle
e^{-\beta(U_{\rm sol}-U_{\rm min})}
\right\rangle ,
\]
by sampling the springs-only reference and evaluating each configuration with
the full TIP4P/2005 potential. This gives
\(\Delta A_1=-62.68\) kJ mol\(^{-1}\), including a
\(+0.39\) kJ mol\(^{-1}\) fluctuation contribution. The solid Helmholtz free
energy is therefore
\[
A_{\rm sol}
=
A_{\rm Einstein}^{\rm id}
+
\Delta A_1
-
W_{\rm spring}
=
34.88-62.68-29.12
=
-56.92~{\rm kJ\,mol}^{-1}.
\]

\paragraph*{Crystal entropy.}
The average energy of the rigid crystal is
\[
\langle E\rangle
=
\langle U\rangle_{\lambda=0}
+
3RT,
\]
where the \(3RT\) term is the kinetic energy of the six rigid degrees of
freedom. With
\(\langle U\rangle_{\lambda=0}=-56.06\) kJ mol\(^{-1}\), this gives
\(\langle E\rangle=-49.77\) kJ mol\(^{-1}\). Including the Pauling residual
entropy of proton disorder,
\(S_{\rm Pauling}=R\ln(3/2)=3.37\) J mol\(^{-1}\) K\(^{-1}\), the ice entropy at
252 K is
\[
S_{\rm ice}(252~{\rm K})
=
\frac{\langle E\rangle-A_{\rm sol}}{T}
+
S_{\rm Pauling}
=
31.7~{\rm J\,mol}^{-1}{\rm K}^{-1}.
\]

\paragraph*{Bridge to the liquid.}
The liquid entropy at 298 K is obtained by adding the model melting entropy and
the heat-capacity integral,
\[
S_{\rm liq}(298)
=
S_{\rm ice}(252)
+
\Delta S_{\rm melt}
+
\int_{252}^{298}\frac{C_p}{T}\,dT .
\]
Using
\(\Delta S_{\rm melt}=19.25\) J mol\(^{-1}\) K\(^{-1}\) and
\(\int C_p/T\,dT=16.55\) J mol\(^{-1}\) K\(^{-1}\), we obtain
\[
S_{\rm liq}(298)
=
31.7+19.25+16.55
=
67.5~{\rm J\,mol}^{-1}{\rm K}^{-1}.
\]
We report this as \(67\pm5\) J mol\(^{-1}\) K\(^{-1}\). The uncertainty is
dominated by the subtraction of large free-energy terms in the crystal entropy
and by the melting/heat-capacity bridge. The value lies above the FEP
self-solvation entropy and below the experimental entropy, and is consistent with the FEP anchor within its uncertainty.

%==============================================================
\section{Sensitivity analysis}
\label{app:filter}
%==============================================================
The cage entropy contains two controls beyond the dimensional prefactor:
a spectral filter that restricts the correction to the cage band, and a
fluidicity gate that suppresses the cage in the crystalline limit. They play
different roles. The filter is a genuine spectral prescription and carries a
measurable systematic uncertainty; the gate is a robust phase switch that is
inactive for the homogeneous liquids studied here.

\paragraph{Friction cutoff.}
In practice $\gamma=\tilde K(0)$ is obtained by integrating the inverted memory
kernel over its numerical support. For a cleanly decaying kernel, this support
ends at the main lobe. A smooth spurious tail, however, can evade the usual
noise and instability guards and overestimate $\gamma$, thereby inflating both
the Markovian gas and the cage residual. We therefore also evaluate $\gamma$ at
a main-lobe cutoff, defined as the first lag for which
$|K(t)|<0.02\,|K(0)|$. If the automatically selected support more than doubles
the friction relative to this main-lobe value, the tail is rejected and the
main-lobe cutoff is used.

This safeguard affected SPC/E water, for which the automatic support extended
to $\sim1$~ps and inflated $\gamma$ by roughly a factor of four. The main-lobe
cutoff reduced the translational cage entropy from
$2.7$ to $1.4$~J\,mol$^{-1}$K$^{-1}$, consistent with TIP4P/2005. TIP4P/2005 has
a cleanly decaying kernel and is unaffected.

\paragraph{Spectral filter.}
The cage correction is multiplied by the low-pass envelope
\[
1-w(\nu),\qquad
w(\nu)=\frac{\nu^2}{\nu^2+\nu_c^2},
\]
with
\[
\nu_c=\frac{\Omega_0}{2\pi c}.
\]
Here $\Omega_0$ is the Einstein frequency obtained from the second moment of the
total density of states. The filter confines the entropy reweighting to the
low-frequency, diffusively coupled part of the memory band, where the
hard-sphere and harmonic references differ most. It also suppresses the
high-frequency tail of the memory excess, which is more harmonic-like and would
otherwise be double-counted as cage entropy. The zero-frequency condition is
imposed independently: since $F_K(0)=F_M(0)=1/\gamma$, the cage spectrum itself
vanishes at $\nu=0$.

\paragraph{Filter-induced onset suppression.}
The same filter shapes the weak-memory limit of the efficiency trend of the
main text: for single-timescale kernels the cage rises as
$(\gamma/\Omega_0)^3$, one power from the memory amplitude and two from the
filter, so weakly non-Markovian systems draw almost
no cage; this is the mechanism behind the strong under-correction on the
Lennard-Jones flank.

\paragraph{Why use the Einstein cutoff.}
A natural alternative is the friction corner,
\[
\nu_c=\frac{\gamma}{2\pi c},
\]
which defines the filter from the Markovian friction rather than from the DoS
second moment. This choice performs slightly better for the Lennard-Jones grid,
but it does not transfer. For the Sutton--Chen metals it gives substantially
larger errors than the Einstein cutoff, and in water it admits too much of the
rotational memory band because $\gamma_{\rm rot}$ is much larger than
$\Omega_0$. We therefore use the Einstein frequency as the transferable
gas--cage discriminator. This choice is deliberately not optimized for the weakly
non-Markovian LJ liquids; it is retained because it avoids over-correcting the
strongly caged systems where the cage entropy is largest.

\paragraph{Cutoff sensitivity.}
The cutoff is not an insensitive numerical convention. At fixed prefactor,
sweeping $\nu_c$ gives approximately
\[
\Delta S_{\rm cage}\propto \nu_c^{0.79},
\]
with a change of order $50\%$ over
$0.5$--$2$ times the nominal cutoff. We therefore treat
$\nu_c=\Omega_0/2\pi c$ as a fixed physical prescription, not as a fitted
parameter. The associated cutoff dependence is a systematic uncertainty of the
present closure.

\paragraph{Fluidicity gate.}
The fluidicity gate
\[
g(f)=\frac{f^2}{f^2+f_0^2},\qquad f_0=0.01,
\]
enforces the crystalline limit. Without this gate, a crystal would still have a
large difference between its phonon spectrum and a Markovian Lorentzian, and the
memory excess would generate a spurious cage entropy. The gate removes this
artifact by forcing $\Delta S_{\rm cage}\to0$ as $f\to0$, so that 3PT reduces to
the Debye harmonic crystal.

Unlike the spectral cutoff, the gate is insensitive to its numerical parameter
over the present data set. Homogeneous fluids have fluidicities well above the
gate scale, whereas crystalline states lie orders of magnitude below it. As a
result, any $f_0$ chosen in the empty gap between these regimes gives the same
liquid entropies to within the reported uncertainty. The gate is therefore a
limit-enforcing switch, not a fitted entropy parameter.

\begin{figure}[tbp]
\includegraphics[width=\columnwidth]{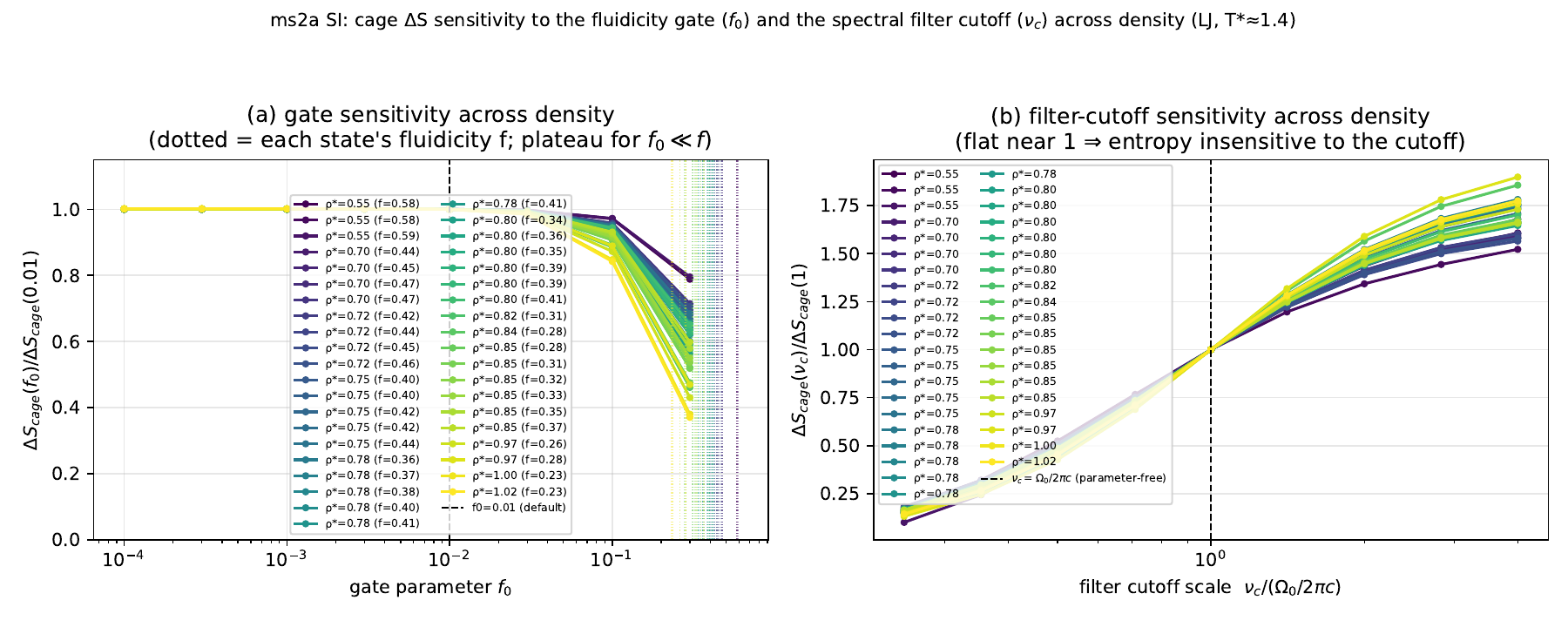}
\caption{\label{fig:sicage}Sensitivity of the cage entropy across the LJ liquid grid (8-fs
trajectories) to (a)~the fluidicity-gate parameter $f_0$ and (b)~the spectral-filter cutoff scale
$\nu_c/(\Omega_0/2\pi c)$. The gate is flat for $f_0$ well below each state's fluidicity (dotted),
so $\Delta S_{\rm cage}$ is independent of $f_0$ in the empty fluidicity gap; the filter cutoff is
not flat ($\Delta S_{\rm cage}\propto\nu_c^{0.79}$, $\sim\!50\%$ over $0.5$--$2\times$), and is fixed
a priori at the Einstein frequency $\nu_c=\Omega_0/2\pi c$ (dashed).}
\end{figure}

\paragraph{Validity of the clip asymmetry.}
The cage retains only the positive lobe of the memory excess $\FK-\FM$; the
compensating negative lobe (a zero-sum partner at the response level) is left in the harmonic solid. To quantify the validity of this prescription, we note that if the negative lobe were instead
reweighted with the same $(1-w)(W_g-W_s)$ envelope, the cage correction would
shrink by $25$--$37\%$ across the LJ liquid grid
($|\Delta S^{-}|/\Delta S^{+}=0.27$, $0.37$, and $0.25$ at
$(\rho^*,T^*)=(0.85,0.9)$, $(0.70,1.1)$, and $(1.00,1.4)$, respectively), and
by comparable fractions ($22$--$41\%$) for analytic one- and two-timescale
model kernels. We retain the clip because the negative lobe lies in the
low-frequency diffusive band whose spectral weight is already assigned to the
gas at its own hard-sphere reference; reweighting it a second time would
double-count the diffusive entropy. The clip and the spectral cutoff are thus both fixed physically motivated constraints, not empirical fit.

\paragraph{Robustness of the Inversion-scheme.}
The time-domain Volterra inversion itself is a fixed prescription, and the friction-support
rule above is its only tuned safeguard. To test this, we re-derived $\gamma$ from the same
VACFs with various alternative schemes, including a first-order finite-difference recursion and a
frequency-domain quotient $\tilde K=(c(0)-i\omega\hat c)/\hat c$ with a small
regularizer. All these attempts shifted $\gamma$ (and the memory area $\mathcal{M}$ with it)
by up to a factor of $\sim\!2$, system-dependently. Likewise, states for which
the automatic support is retained would now carry an error of up to
$\sim\!1.6\times$ in $\gamma$. The inversion scheme is therefore part of the method definition, on the same footing as the spectral cutoff and the clip; the resulting ambiguity is reflected in the state-to-state scatter about the efficiency trend reported in the main text.

%==============================================================
\section{Complete LJ \texorpdfstring{$(\rho^*,T^*)$}{(rho,T)} grid}\label{app:grid}
%==============================================================
Table~\ref{tab:s1} gives all six entropies ($S^*$ in $\kB$/atom) computed self-consistently from the
same trajectories (FSC off; $N=512$ for the fluid states, $N=500$ FCC for the $\rho^*\!\ge\!1.05$
solids; R2PT is undefined for the near-solid $\rho^*>1.02$ rows). The reference is the MBWR fluid EOS for the gas, SC-fluid and liquid rows and the van der
Hoef solid EOS~\cite{vdHoef2000} for the solid rows; $\Delta_{\rm 3PT}=S^*_{\rm 3PT}-S^*_{\rm ref}$.
The 43 liquid rows ($0.70\le\rho^*\le1.02$) give the headline RMS deviations of $0.16$ (3PT),
$0.14$ (Lin-2003) and $0.33$ (rig-HS) quoted in Sec.~\ref{M-sec:lj}. The gas rows
($\rho^*\!\le\!0.25$) carry negligible cage weight; the solid rows ($\rho^*\!\ge\!1.05$) have
$f\to0$ (Lin-2003 $\equiv$ rig-HS there) where the fluidicity gate $g(f)$
(Sec.~\ref{M-sec:cageentropy}) switches the cage off, so 3PT also returns to rig-HS (the
Debye--Einstein crystal). Neither region is part of the liquid-cage domain that defines the headline
RMS. $(\rho^*,T^*)=(0.4,0.9)$ ($^\dagger$) lies inside the liquid--vapor coexistence envelope, where
the single-phase MBWR reference does not apply.

\begin{longtable}{c c l c c c c c c c}
\caption{\label{tab:s1}Complete LJ $(\rho^*,T^*)$ grid; $S^*$ in $\kB$/atom. R2PT is undefined for the near-solid $\rho^*>1.02$ rows (---).}\\
\hline\hline
$\rho^*$ & $T^*$ & region & ref & rig-HS & Lin-2003 & Desj. & R2PT & 3PT & $\Delta_{\rm 3PT}$ \\
\hline
\endfirsthead
\multicolumn{10}{l}{\footnotesize Table~\ref{tab:s1} (continued)}\\
\hline\hline
$\rho^*$ & $T^*$ & region & ref & rig-HS & Lin-2003 & Desj. & R2PT & 3PT & $\Delta_{\rm 3PT}$ \\
\hline
\endhead
\hline
\endfoot
\hline\hline
\endlastfoot
0.050 & 1.400 & gas & 13.671 & 13.634 & 13.752 & 13.734 & 13.600 & 13.646 & $-$0.025 \\
0.050 & 1.600 & gas & 13.885 & 14.049 & 14.172 & 14.149 & 14.043 & 14.000 & $+$0.115 \\
0.050 & 1.800 & gas & 14.071 & 14.060 & 14.161 & 14.141 & 14.022 & 14.052 & $-$0.019 \\
0.050 & 2.000 & gas & 14.235 & 14.300 & 14.401 & 14.382 & 14.281 & 14.278 & $+$0.043 \\
0.100 & 1.250 & gas & 12.636 & 12.428 & 12.626 & 12.603 & 12.401 & 12.431 & $-$0.205 \\
0.150 & 1.300 & gas & 12.154 & 11.911 & 12.154 & 12.130 & 11.912 & 11.906 & $-$0.248 \\
0.200 & 1.400 & gas & 11.873 & 11.611 & 11.880 & 11.859 & 11.623 & 11.604 & $-$0.269 \\
0.200 & 1.600 & gas & 12.125 & 11.943 & 12.213 & 12.187 & 12.021 & 11.930 & $-$0.195 \\
0.200 & 1.800 & gas & 12.334 & 12.270 & 12.541 & 12.514 & 12.365 & 12.243 & $-$0.091 \\
0.200 & 2.000 & gas & 12.515 & 12.443 & 12.703 & 12.676 & 12.525 & 12.420 & $-$0.095 \\
0.250 & 1.300 & SC-fluid & 11.363 & 11.053 & 11.378 & 11.359 & 11.165 & 11.045 & $-$0.318 \\
0.300 & 1.300 & SC-fluid & 11.051 & 10.717 & 11.060 & 11.047 & 10.815 & 10.714 & $-$0.337 \\
0.320 & 1.320 & SC-fluid & 10.969 & 10.641 & 10.980 & 10.969 & 10.710 & 10.643 & $-$0.326 \\
0.350 & 1.350 & SC-fluid & 10.851 & 10.528 & 10.878 & 10.869 & 10.608 & 10.524 & $-$0.327 \\
0.400 & 0.900$^\dagger$ & SC-fluid & 9.512 & 8.445 & 8.896 & 8.909 & 8.612 & 8.526 & $-$0.986 \\
0.400 & 1.400 & SC-fluid & 10.661 & 10.394 & 10.796 & 10.788 & 10.601 & 10.380 & $-$0.281 \\
0.400 & 1.600 & SC-fluid & 10.920 & 10.658 & 11.036 & 11.028 & 10.823 & 10.653 & $-$0.267 \\
0.400 & 1.800 & SC-fluid & 11.138 & 10.938 & 11.323 & 11.314 & 11.138 & 10.924 & $-$0.214 \\
0.400 & 2.000 & SC-fluid & 11.327 & 11.170 & 11.550 & 11.541 & 11.371 & 11.150 & $-$0.177 \\
0.550 & 1.400 & SC-fluid & 9.873 & 9.574 & 10.000 & 10.009 & 9.756 & 9.615 & $-$0.258 \\
0.550 & 1.600 & SC-fluid & 10.129 & 9.851 & 10.271 & 10.279 & 10.020 & 9.891 & $-$0.238 \\
0.550 & 1.800 & SC-fluid & 10.351 & 10.097 & 10.531 & 10.537 & 10.341 & 10.127 & $-$0.224 \\
0.700 & 1.000 & liquid & 8.281 & 7.917 & 8.412 & 8.432 & 8.219 & 8.041 & $-$0.240 \\
0.700 & 1.100 & liquid & 8.487 & 8.124 & 8.617 & 8.637 & 8.427 & 8.247 & $-$0.240 \\
0.700 & 1.250 & liquid & 8.755 & 8.446 & 8.915 & 8.937 & 8.683 & 8.571 & $-$0.184 \\
0.700 & 1.400 & liquid & 8.990 & 8.675 & 9.161 & 9.181 & 8.981 & 8.792 & $-$0.198 \\
0.720 & 1.000 & liquid & 8.145 & 7.766 & 8.260 & 8.281 & 8.063 & 7.896 & $-$0.249 \\
0.720 & 1.100 & liquid & 8.353 & 7.998 & 8.491 & 8.512 & 8.301 & 8.132 & $-$0.221 \\
0.720 & 1.250 & liquid & 8.625 & 8.288 & 8.770 & 8.792 & 8.565 & 8.436 & $-$0.189 \\
0.720 & 1.400 & liquid & 8.863 & 8.552 & 9.029 & 9.051 & 8.825 & 8.681 & $-$0.182 \\
0.750 & 0.900 & liquid & 7.688 & 7.314 & 7.805 & 7.826 & 7.594 & 7.458 & $-$0.230 \\
0.750 & 1.000 & liquid & 7.934 & 7.563 & 8.059 & 8.080 & 7.867 & 7.705 & $-$0.229 \\
0.750 & 1.100 & liquid & 8.148 & 7.797 & 8.289 & 8.311 & 8.094 & 7.927 & $-$0.221 \\
0.750 & 1.250 & liquid & 8.426 & 8.086 & 8.580 & 8.601 & 8.399 & 8.214 & $-$0.212 \\
0.750 & 1.400 & liquid & 8.670 & 8.349 & 8.837 & 8.860 & 8.651 & 8.477 & $-$0.193 \\
0.780 & 0.850 & liquid & 7.318 & 6.966 & 7.457 & 7.476 & 7.253 & 7.120 & $-$0.198 \\
0.780 & 0.900 & liquid & 7.461 & 7.101 & 7.593 & 7.613 & 7.391 & 7.275 & $-$0.186 \\
0.780 & 1.000 & liquid & 7.716 & 7.349 & 7.841 & 7.862 & 7.643 & 7.514 & $-$0.202 \\
0.780 & 1.100 & liquid & 7.937 & 7.589 & 8.078 & 8.099 & 7.874 & 7.752 & $-$0.185 \\
0.780 & 1.250 & liquid & 8.222 & 7.884 & 8.377 & 8.399 & 8.193 & 8.017 & $-$0.205 \\
0.780 & 1.400 & liquid & 8.471 & 8.156 & 8.650 & 8.672 & 8.478 & 8.288 & $-$0.183 \\
0.800 & 0.850 & liquid & 7.157 & 6.812 & 7.297 & 7.316 & 7.090 & 6.998 & $-$0.159 \\
0.800 & 0.900 & liquid & 7.305 & 6.974 & 7.455 & 7.476 & 7.229 & 7.167 & $-$0.138 \\
0.800 & 1.000 & liquid & 7.567 & 7.216 & 7.703 & 7.723 & 7.508 & 7.381 & $-$0.186 \\
0.800 & 1.100 & liquid & 7.792 & 7.462 & 7.943 & 7.967 & 7.726 & 7.618 & $-$0.174 \\
0.800 & 1.250 & liquid & 8.083 & 7.752 & 8.243 & 8.265 & 8.053 & 7.903 & $-$0.180 \\
0.800 & 1.400 & liquid & 8.337 & 8.024 & 8.515 & 8.538 & 8.332 & 8.177 & $-$0.160 \\
0.820 & 0.800 & liquid & 6.831 & 6.506 & 6.981 & 6.998 & 6.766 & 6.712 & $-$0.119 \\
0.840 & 0.720 & liquid & 6.402 & 6.069 & 6.525 & 6.538 & 6.286 & 6.233 & $-$0.169 \\
0.850 & 0.750 & liquid & 6.402 & 6.098 & 6.555 & 6.569 & 6.309 & 6.362 & $-$0.040 \\
0.850 & 0.900 & liquid & 6.898 & 6.569 & 7.037 & 7.055 & 6.811 & 6.805 & $-$0.093 \\
0.850 & 1.000 & liquid & 7.180 & 6.828 & 7.304 & 7.323 & 7.098 & 7.045 & $-$0.135 \\
0.850 & 1.100 & liquid & 7.420 & 7.068 & 7.547 & 7.566 & 7.347 & 7.279 & $-$0.141 \\
0.850 & 1.250 & liquid & 7.727 & 7.392 & 7.876 & 7.898 & 7.686 & 7.589 & $-$0.138 \\
0.850 & 1.400 & liquid & 7.992 & 7.677 & 8.165 & 8.187 & 7.983 & 7.863 & $-$0.129 \\
0.900 & 0.750 & liquid & 5.952 & 5.663 & 6.072 & 6.078 & 5.787 & 5.907 & $-$0.045 \\
0.900 & 0.800 & liquid & 6.120 & 5.839 & 6.263 & 6.270 & 5.998 & 6.094 & $-$0.026 \\
0.950 & 0.800 & liquid & 5.651 & 5.404 & 5.769 & 5.772 & 5.434 & 5.641 & $-$0.010 \\
0.950 & 0.850 & liquid & 5.833 & 5.573 & 5.949 & 5.952 & 5.631 & 5.814 & $-$0.019 \\
0.975 & 1.400 & liquid & 7.078 & 6.764 & 7.205 & 7.216 & 6.986 & 7.037 & $-$0.041 \\
0.975 & 1.600 & liquid & 7.430 & 7.127 & 7.582 & 7.595 & 7.384 & 7.300 & $-$0.130 \\
0.980 & 0.900 & liquid & 5.727 & 5.493 & 5.851 & 5.853 & 5.512 & 5.721 & $-$0.006 \\
1.000 & 0.950 & liquid & 5.714 & 5.483 & 5.822 & 5.822 & 5.461 & 5.686 & $-$0.028 \\
1.000 & 1.400 & liquid & 6.886 & 6.587 & 7.008 & 7.017 & 6.774 & 6.862 & $-$0.024 \\
1.020 & 1.500 & liquid & 6.922 & 6.628 & 7.045 & 7.055 & 6.814 & 6.904 & $-$0.018 \\
1.050 & 0.800 & solid & 4.314 & 4.322 & 4.326 & 4.328 & --- & 4.314 & $-$0.000 \\
1.100 & 1.000 & solid & 4.584 & 4.578 & 4.581 & 4.582 & --- & 4.547 & $-$0.037 \\
1.100 & 1.600 & solid & 5.897 & 5.790 & 5.793 & 5.795 & --- & 5.773 & $-$0.124 \\
1.100 & 1.800 & solid & 6.226 & 6.110 & 6.114 & 6.116 & --- & 6.086 & $-$0.140 \\
1.150 & 1.600 & solid & 5.570 & 5.514 & 5.516 & 5.518 & --- & 5.498 & $-$0.072 \\
1.200 & 2.000 & solid & 5.886 & 5.819 & 5.822 & 5.824 & --- & 5.789 & $-$0.097 \\
\end{longtable} 

\noindent{\footnotesize $^\dagger$Inside the LJ liquid--vapor coexistence envelope; the
single-phase MBWR reference does not apply (every method deviates strongly, rig-HS $-1.06$).
Reported for completeness; not a homogeneous-fluid test point. Solid rows ($\rho^*\ge1.05$) use
$N=500$ FCC trajectories with the van der Hoef solid EOS as reference. With the fluidicity gate
$g(f)$ active ($f\lesssim4\times10^{-4}$, so $g<10^{-2}$) the cage switches off and 3PT $=$ rig-HS,
recovering the Debye--Einstein crystal accurate to $\Delta=-0.01$ to $-0.15\,\kB$; without the gate
the bare memory excess would over-correct these rows to $\Delta=-0.14$ to $-0.46$
(Sec.~\ref{M-sec:cageentropy}).}

\bibliography{refs}

% Alt Text is accessibility metadata, not a document section: suppress its TOC entry
% (REVTeX writes even starred \section* to the .toc) while keeping section styling.
\begingroup
\renewcommand{\addcontentsline}[3]{}%
\section*{Alt Text}
\endgroup

\textbf{Figure~\ref{fig:d4prefactor}.} Two-panel test of the cage prefactor in four dimensions. The left panel plots, against the memory-area metric, the non-memory (soft-mode anharmonic) entropy per mode that each candidate prefactor (1/4, 1/3, 1/2) would require to place the cage efficiency on the three-dimensional universal curve, for 22 four-dimensional Lennard-Jones fluid states; the curve for p=1/4 is flat and lands within the shaded per-mode anharmonic band, p=1/3 lies just below it, and p=1/2 falls below the band with a memory-area-dependent slope and is excluded. The right panel shows the distribution of that required residual for each prefactor as a box plot, confirming that p=1/4 is the flat, in-band choice.

\textbf{Figure~\ref{fig:sicage}.} Two-panel sensitivity plot for the cage entropy across the Lennard-Jones liquid grid. The left panel shows the cage entropy is flat against the fluidicity-gate parameter over a wide range, so the gate is effectively parameter-free; the right panel shows the cage entropy rises monotonically with the spectral-filter cutoff (no plateau), so the cutoff is a genuine choice, fixed a priori at the Einstein frequency.

\textbf{Table~\ref{tab:fep}.} Table assembling the FEP-based standard molar entropy of TIP4P/2005 and SPC/E water at 298 K. It combines ideal translational and rotational entropies, intermolecular energy, and self-solvation free energy to obtain the final entropy.

\textbf{Table~\ref{tab:s1}.} Long table of Lennard-Jones entropies over the full density–temperature grid. It compares reference values with several 2PT-family methods and reports the 3PT deviation across gas, supercritical-fluid, liquid, and solid regimes.